\documentclass[english,twocolumn,prd,nofootinbib]{revtex4}
\usepackage{amsmath}
\usepackage{amssymb}
\usepackage{amsfonts}

\begin{document}

\title{Gravitational dynamics in $s+1+1$ dimensions II. Hamiltonian theory}
\author{Zolt\'{a}n Kov\'{a}cs$^{1,2}$, L\'{a}szl\'{o} \'{A}. Gergely$^{2}$}
\affiliation{$^{1}$Max-Planck-Institut f{\"{u}}r Radioastronomie, Auf dem H{\"{u}}gel 69,
53121 Bonn, Germany\\
$^{2}$Departments of Theoretical and Experimental Physics, University of
Szeged, D\'{o}m t\'{e}r 9, Szeged 6720, Hungary}

\begin{abstract}
We develop a Hamiltonian formalism of brane-world gravity, which singles out
two preferred, mutually orthogonal directions. One is a unit twist-free
field of spatial vectors with integral lines intersecting perpendicularly
the brane. The other is a temporal vector field with respect to which we
perform the Arnowitt-Deser-Misner decomposition of the Einstein-Hilbert
Lagrangian. The gravitational variables arise from the projections of the
spatial metric and their canonically conjugated momenta as tensorial,
vectorial and scalar quantities defined on the family of hypersurfaces
containing the brane. They represent the gravitons, a gravi-photon and a
gravi-scalar, respectively. From the action we derive the canonical
evolution equations and the constraints for these gravitational degrees of
freedom both on the brane and outside it. By integrating across the brane,
the dynamics also generates the tensorial and scalar projection of the
Lanczos equation. The vectorial projection of the Lanczos equation arises in
a similar way from the diffeomorphism constraint. Both the graviton and the
gravi-scalar are continuous across the brane, however the momentum of the
gravi-vector has a jump, related to the energy transport (heat flow) on the
brane.
\end{abstract}

\maketitle

\section{Introduction}

The Hamiltonian theory of general relativity is based on the
Arnowitt-Deser-Misner (ADM) decomposition \cite{ADM}, i.e. slicings of the
four dimensional (4d) space-time manifold with a family of spacelike
three-dimensional (3d) hypersurfaces.

The slicing method with respect to a timelike hypersurface (the brane) has
been also proven efficient in brane world models which interpret our
traditional 4d space-time as a brane embedded in a $5$-dimensional (5d)
manifold, the bulk (for a review see \cite{MaartensLR}). In the $4+1$
decomposition of the bulk, various geometrical projections of the 5d gravity
on the brane were used together with the Lanczos-Sen-Darmois-Israel (LSDI)
matching conditions \cite{Lanczos}-\cite{Israel} in order to derive the
effective Einstein equation on the brane \cite{SMS}-\cite{Decomp}. In these
approaches the 4d space-time manifold was treated covariantly. In the
covariant approach however it is not obvious how to describe time evolution,
and introduce the very concept of the (many-fingered) time function, which
is the first step in developing a Hamiltonian formulation of the brane world
scenario.

In a previous paper \cite{Paper1} (hereafter quoted as Paper \textbf{I}), we
have discussed a two-fold slicing of the bulk, both with respect to a
timelike and a spacelike foliation. One of them was meant to contain the
brane, the other the constant time slices $\Sigma _{t}$ (referring to $t$%
=const) of the bulk, necessary for Hamiltonian evolution. The result was a $%
\left( 3+1+1\right) $-decomposition of the bulk, which resulted in a
break-up of the brane into space and time. The formalism was given in fact
for the more generic case of $s+1+1$ dimensions, therefore it is applicable
for general relativistic situations as well, whenever any space-like
direction should be singled out for one reason or another.

In this framework the gravitational degrees of freedom are represented by
new variables. These are the spatial metric $g_{ab}$ on the brane, the shift
vector $M^{a}$ and lapse function $M$ \textit{associated with the brane
normal}; together with the tensorial, vectorial and scalar projections of
the extrinsic curvature associated with the temporal normal. These can be
thought of as gravi-tensorial, gravi-vectorial and gravi-scalar degrees of
freedom, the extrinsic curvatures representing their generalized velocities.

In Paper \textbf{I}\ we have given the decomposition of all geometric
quantities with respect to the two normals, together with the evolution
equations of the gravi-tensorial, gravi-vectorial and gravi-scalar degrees
of freedom in the velocity phase-space. A set of (Lagrangian-type)
constraint equations on these variables were also given. The latter should
be imposed on any set of initial data on a spacelike section of the brane.

However Paper \textbf{I}\ did not contain any reference to canonical momenta
and the phase-space of gravitational degrees of freedom. Neither did it
address the variational principle leading to the Hamiltonian evolution
equations, together with the super-Hamiltonian and super-momentum
(diffeomorphism) constraints of brane-world gravity.

It is the purpose of this paper to discuss these topics, together with the
regularization issues raised by the presence of the distributional matter
sources on the brane. In Sec. II we give a brief, somewhat technical account
of those results presented in Paper \textbf{I} which are needed for our
present considerations.

Sec. III contains the variational principle for vacuum gravity, with
concomitant ADM and brane-bulk decompositions. We define here the momenta
associated with the gravi-tensorial, gravi-vectorial and gravi-scalar
degrees of freedom and perform the Legendre-transformation. Extremizing the
action with respect to the gravitational phase-space variables give the
Hamiltonian evolution equations in terms of variables defined on the $t=$%
const space sections of the brane. Extremizing with respect to the lapse $%
N^{a}$ and shift $N$ (both related to the temporal evolution) gives the
super-Hamiltonian and super-momentum constraints of brane-world gravity.
Both the shift $N^{a}$ and super-momentum $\mathcal{H}_{a}^{G}$ have only $s$
components, as the off-brane component of the shift is suppressed in order
to obey Frobenius' theorem. i.e. to let the brane exist as a hypersurface.%
\footnote{%
While in the $s+1$ ADM decomposition the congruence $\partial /\partial t$
generating time-evolution should be twist-free in order $\Sigma _{t}$ to
exist, there is no such condition, that an $\left( s-1\right) $-parameter
subset of the congruence should form a temporal hypersurface. It is exactly
this condition, which should be obeyed in brane-worlds, leading in the
simplest case to the suppression of the $\left( s+1\right) ^{\text{th}}$
component of the shift.}

Matter sources both in the bulk and on the brane are discussed in Sec. IV.
We show here the following commutativity property: the $\left( s+1+1\right) $%
-decomposition of the energy-momentum tensor obtained by a covariant
variation leads to the same result as the non-covariant variation with
respect to the gravitational variables arising from the $\left( s+1+1\right) 
$-decomposition. We also give here the Hamiltonian evolution in the presence
of matter. Similarly as the evolution equations in the velocity phase-space
given in Paper \textbf{I}, these Hamiltonian evolution equations contain
distributional type sources.

We remedy this situation in Sec. V by a regularization procedure across the
brane. We consider an interval $\left( -\chi _{0},~\chi _{0}\right) $ across
the brane (lying at $\chi =0$) and integrate the super-Hamiltonian and
super-momentum constraints of the total system (gravity + matter). We also
integrate the Hamiltonian evolution equation of the gravi-tensorial
momentum. Finally we take the thin brane limit $\chi _{0}\rightarrow 0$.
This method gives the scalar, vectorial and tensorial projections (with
respect to $\Sigma _{t}$) of the Lanczos equation.

Sec. VI contains the concluding remarks. We include three Appendices, the
first containing the $\left( s+1+1\right) $-decompositions of various
quantities needed throughout the paper. In the second Appendix we schetch
the derivation of the projections of the Lanczos equation in the velocity
phase-space. The third Appendix proves another commutativity property. While
in the main paper we obtain the results by an $s+1+1$ ADM\ decomposition
prior to applying the variational principle; the same results can be
regained by first performing an $\left( s+1\right) +1$ ADM reduction, then
extremizing the action and finally by a further$\left( s+1\right) $-split.

\textit{Notation.---} All quantities defined on the full $\left(s+2\right)$%
-dimensional space-time and on the $\left(s+1\right)$-dimensional brane
carry a distinguishing tilde and the prefix $\left(s+1\right)$,
respectively. A hat distinguishes quantities defined on the spatial $%
\left(s+1\right)$-leaves. Quantities defined on the spatial sections of the
brane carry no distinctive mark. For example, the metric 2-forms are denoted 
$\widetilde{g}$, $\widehat{g}$ and $g$, respectively, while the
corresponding metric-compatible connections are $\widetilde{\nabla}$, $%
\widehat{D}$ and $D$. Latin indices represent abstract indices running from $%
0$ to $(s+1)$. Vector fields in Lie-derivatives are represented by boldface
characters. For example $\pounds _{\mathbf{V}}T$ denotes the Lie derivative
along the integral lines of the vector field $V^{a}$. A dot denotes a
derivative with respect to $t$, such that $\dot{T}=\left(\partial/\partial
t\right)T=\pounds _{\partial/\partial t}T$.

\section{The $\left( s+1+1\right) $-decomposition of space-time}

In order to develop the Hamiltonian theory of brane-world gravity from the
viewpoint of an observer on the brane, a double decomposition of the bulk is
necessary. We have to separate both the normal to the brane $l^{a}$ and the
temporal normal $n^{a}$, with respect to which the ADM decomposition will be
carried out. These vector fields obey $n^{a}n_{a}=-1$, $l^{a}l_{a}=1$ and $%
n^{a}l_{a}=0$. We introduce a coordinate $\chi$ transverse to the brane and
we foliate the $\left(s+1+1\right)$-dimensional space-time by $%
\left(s+1\right)$-dimensional leaves of constant $\chi$, the brane being at $%
\chi=0$. The \textit{off-brane evolution} is defined along the integral
lines of%
\begin{equation}
\left(\frac{\partial}{\partial\chi}\right)^{a}=M^{a}+Ml^{a}\ ,
\label{off-brane}
\end{equation}
where $M^{a}$ and $M$ represent the shift vector and lapse function
associated with the off-brane evolution. They characterize the off-brane
sector of gravity.

The temporal evolution is along the integral lines of%
\begin{equation}
\left( \frac{\partial }{\partial t}\right) ^{a}=Nn^{a}+N^{a}\ ,\qquad
\label{temporal}
\end{equation}%
where $N^{a}$ and $N$ are the familiar shift vector and lapse function
related to the foliation of the $\left( s+1+1\right) $-dimensional
space-time by the $\left( s+1\right) $-dimensional spatial leaves $t=$const.
The coordinate $t$ represents time. Note that neither $\left( \partial
/\partial \chi \right) ^{a}$ has a component along $n^{a}$, nor $\left(
\partial /\partial t\right) ^{a}$ along $l^{a}$. The first condition
guarantees that $\left( \partial /\partial \chi \right) ^{a}$ generates a
purely spatial displacement. The second one is the simplest gauge-choice,
which ensures that the Fr\"{o}benius theorem holds, thus $l^{a}$ is
hypersurface-forming (see Appendix C of Paper \textbf{I}).

The metric $\widetilde{g}_{ab}$ of the $\left( s+1+1\right) $-dimensional
space-time, decomposed with respect to $l^{a}$ and $n^{a}~$is%
\begin{equation}
\widetilde{g}_{ab}=g_{ab}-n_{a}n_{b}+l_{a}l_{b}\ ,  \label{tildeg0}
\end{equation}%
where $g_{ab}$ is the induced metric on the $s$-dimensional leaves $\Sigma
_{t\chi }$, which represent the intersection of the $\chi $=const and $t$%
=const leaves. The components of $\left( s+2\right) $-dimensional metric in
the coordinate system adapted to the coordinates $(t,\chi ,x^{a})$ can be
written as%
\begin{gather}
ds^{2}=-(N_{a}N^{a}-N^{2})dt^{2}+2N_{a}dtdx^{a}+N_{a}M^{a}dtd\chi  \notag \\
+g_{ab}dx^{a}dx^{b}+2M_{a}dx^{a}d\chi +(M_{a}M^{a}+M^{2})d\chi ^{2}\;.
\label{ds2}
\end{gather}%
Then, as discussed in Paper \textbf{I}, the gravitational sector is
described by $\{g_{ab},M^{a},M,N^{a},N\}$ obeying $%
g_{ab}n^{a}=g_{ab}l^{a}=M^{a}n_{a}=M^{a}l_{a}=N^{a}n_{a}=N^{a}l_{a}=0$. The
above set has one variable less then the number of variables contained in $%
\widetilde{g}_{ab}$. This is because time-evolution is restricted to proceed
along $\chi $=const hypersurfaces. (According to Eq. (\ref{temporal}) there
is no off-brane component of the shift.)

In Paper \textbf{I} we have introduced two types of extrinsic curvatures,
related to the normal vector fields $n^{a}$ and $l^{a}$. Each of them could
be further decomposed with respect to the other normal into tensorial,
vectorial and scalar projections. These give two kinds of second fundamental
forms of the leaves $\Sigma_{t\chi}$ 
\begin{equation}
K_{ab}=g^{c}{}_{a}g^{d}{}_{b}\widetilde{\nabla}_{c}n_{d}\ ,\qquad
L_{ab}=g^{c}{}_{a}g^{d}{}_{b}\widetilde{\nabla}_{c}l_{d}\;\;,\ 
\label{extr3}
\end{equation}
two normal fundamental forms 
\begin{equation}
\mathcal{K}_{a}=g^{b}{}_{a}l^{c}\widetilde{\nabla}_{c}n_{b}\ ,\qquad\mathcal{%
L}_{a}=-g^{b}{}_{a}n^{c}\widetilde{\nabla}_{c}l_{b}  \label{calK1}
\end{equation}
and two normal fundamental scalars 
\begin{equation}
\mathcal{K}=l^{a}l^{b}\widetilde{\nabla}_{a}n_{b}\ ,\qquad\mathcal{L}%
=n^{a}n^{b}\widetilde{\nabla}_{a}l_{b}\ .  \label{calK2}
\end{equation}
From the symmetry property of the extrinsic curvatures and $n^{a}l_{a}=0$
the condition: $\mathcal{K}_{a}=\mathcal{L}_{a}\ $follows.

It was shown in Paper \textbf{I} that $L_{ab}$ and $\mathcal{L}$ can be
expressed in terms of $\chi $-derivatives and the covariant derivatives $%
D_{a}$ associated with $g_{ab}$ as 
\begin{subequations}
\begin{eqnarray}
L_{ab} &=&\frac{1}{2M}\left( \frac{\partial g_{ab}}{\partial \chi }%
-2D_{(a}M_{b)}\right) \ , \\
\mathcal{L} &=&-\frac{1}{MN}\left( \frac{\partial N}{\partial \chi }%
-M^{a}D_{a}N\right) \ .
\end{eqnarray}%
Thus, $L_{ab}$ and $\mathcal{L}$ are related to the spatial derivatives of
the gravitational variables. By contrast, the quantities $K_{ab}$, $\mathcal{%
K}_{a}$ and $\mathcal{K}$ give the time evolution of $g_{ab}$, $M^{a}$ and $M
$, respectively: 
\end{subequations}
\begin{subequations}
\label{Ks}
\begin{eqnarray}
\ K_{ab}\! &=&\frac{1}{2N}\left( \frac{\partial g_{ab}}{\partial t}%
-2D_{(a}N_{b)}\right) \ , \\
\mathcal{K}^{a}\! &=&\!\!\!\frac{1}{2MN}\!\left( \!\!\frac{\partial M^{a}}{%
\partial t}\!-\!\frac{\partial N^{a}}{\partial \chi }\!+\!M^{b}\!D_{b}%
\!N^{a}\!\!-\!\!N^{b}\!D_{b}\!M^{a}\!\!\right) \!~, \\
\mathcal{K\!} &=&\frac{1}{MN}\left( \frac{\partial M}{\partial t}%
-N^{a}D_{a}M\right) \ .
\end{eqnarray}%
therefore they are velocity-type variables.

Thus gravitational dynamics in the velocity phase-space can be given in
terms of $\{ g_{ab},M^{a},M;K_{ab},\mathcal{K}^{a},\mathcal{K}\}.$
Time-evolution of $\{ g_{ab},M^{a},M\}$ is expressed by Eqs. (\ref{Ks}),
while equations representing the time-evolution of $\{ K_{ab},\mathcal{K}%
^{a},\mathcal{K}\}$ were given by Eqs. (67) of Paper \textbf{I} . \ 

In the next section we present the Hamiltonian formulation of brane-world
gravity, by passing to the momentum phase-space.

\section{Hamiltonian evolution of vacuum gravity in the bulk}

Vacuum geometrodynamics (without cosmological constant) arises from the
Einstein-Hilbert action 
\end{subequations}
\begin{equation}
S^{G}[\widetilde{g}_{ab}]=\int d^{s+2}x\mathcal{L}^{G}=\int d^{s+2}x\sqrt{-%
\widetilde{g}}\widetilde{R}\;.  \label{actionEH}
\end{equation}%
The $\left( s+1+1\right) $-decomposition of the $(s+2)$-dimensional scalar
curvature and of the metric determinant were derived in Paper \textbf{I} as
Eqs. (61) and (B2), respectively. The vacuum gravitational Lagrangian
density becomes%
\begin{eqnarray}
&&\mathcal{L}^{G}[g_{ab},M^{a},M,K_{ab},\mathcal{K}_{a},\mathcal{K};N^{a},N]=
\notag \\
&&NM\sqrt{g}(R-L_{ab}L^{ab}+L^{2}-2\mathcal{L}L)+2\sqrt{g}D_{a}ND^{a}M 
\notag \\
&&+NM\sqrt{g}(K_{ab}K^{ab}-K^{2}+2\mathcal{K}_{a}\mathcal{K}^{a}-2\mathcal{K}%
K)  \notag \\
&&-2\widetilde{\nabla }_{a}[NM\sqrt{g}(\alpha ^{a}-\lambda
^{a}-Kn^{a}+Ll^{a})]~.  \label{LG}
\end{eqnarray}%
Here $\alpha ^{a}=n^{b}\widetilde{\nabla }_{b}n^{a}=N^{-1}D^{a}N-\mathcal{L}%
l^{a}$ and $\lambda ^{a}=l^{b}\widetilde{\nabla }_{b}l^{a}=-M^{-1}D^{a}M+%
\mathcal{K}n^{b}$ are acceleration-type quantities (the curvatures of the
normal congruences $n^{a}$ and $l^{a}$). By transforming (the double of) the
terms quadratic in the extrinsic curvatures in the Lagrangian density (\ref%
{LG}) into time-derivative terms cf. Eqs. (\ref{Ks}) we obtain the following
advantageous expression: 
\begin{eqnarray}
&&\mathcal{L}^{G}[g_{ab},M^{a},M,K_{ab},\mathcal{K}_{a},\mathcal{K};N^{a},N]=
\notag \\
&&\sqrt{g}M\left[ K^{ab}-\left( K+\mathcal{K}\right) g^{ab}\right] \dot{g}%
_{ab}+2\sqrt{g}\mathcal{K}_{a}\dot{M}^{a}  \notag \\
&&-2\sqrt{g}K\dot{M}-N\mathcal{H}_{\bot }^{G}-N^{a}\mathcal{H}_{a}^{G} 
\notag \\
&&-2\sqrt{g}D_{a}\left[ M(D^{a}N+N_{b}K^{ab})+N\mathcal{L}M^{a}-N_{b}%
\mathcal{K}^{b}M^{a}\right]   \notag \\
&&+2\frac{\partial }{\partial \chi }\left[ \sqrt{g}\left( N\mathcal{L\!}%
-\!N_{a}\mathcal{K}^{a}\right) \right] +2\frac{\partial }{\partial t}\left[ M%
\sqrt{g}(K+\mathcal{K})\right] ~,  \label{LG0}
\end{eqnarray}%
where we have denoted 
\begin{subequations}
\label{constraintsG}
\begin{eqnarray}
\mathcal{H}_{\bot }^{G} &=&-\sqrt{g}[M(R+L^{2}-3L_{ab}L^{ab})  \notag \\
&&-2g^{ab}(\partial /\partial \chi -\pounds _{\mathbf{M}})L_{ab}-2D_{a}D^{a}M
\notag \\
&&+M(K^{2}-K_{ab}K^{ab}-2\mathcal{K}_{a}\mathcal{K}^{a}+2K\mathcal{K})]\;,
\label{constraintsVham} \\
\mathcal{H}_{a}^{G} &=&-\sqrt{g}\{D_{b}[MK^{b}{}_{a}-Mg^{b}{}_{a}(K+\mathcal{%
K})]+KD_{a}M  \notag \\
&&+M\mathcal{K}_{a}L+(\partial /\partial \chi -\pounds _{\mathbf{M}})%
\mathcal{K}_{a}\}\;.  \label{constraintsVdiff}
\end{eqnarray}%
(Dots represent the time-derivatives $\partial /\partial t$). The advantage
of writing the Lagrangian density in the form (\ref{LG0}) is that it
contains explicitly the Liuville form, with the right coefficient to drop
out when performing later on the Legendre transformation.\ The cofactors $%
\mathcal{H}_{\bot }^{G}$ and $\mathcal{H}_{a}^{G}$ of the Lagrange
multipliers $N$ and $N^{a}$ are the super-Hamiltonian and the super-momentum
constraints of vacuum gravity, also obtainable as projections of the
Einstein tensor, cf. Appendix A: 
\end{subequations}
\begin{subequations}
\begin{eqnarray}
\mathcal{H}_{\bot }^{G} &=&-2M\sqrt{g}n^{a}n^{b}\widetilde{G}_{ab}\;, \\
\mathcal{H}_{a}^{G} &=&-2M\sqrt{g}g^{b}{}_{a}n^{c}\widetilde{G}_{bc}\;.
\end{eqnarray}%
Since the $\left( s+1\right) $-th component of the shift was set to zero,
the super-momentum contains only the components corresponding to brane
spatial diffeomorphisms.

Now we define the phase space of the brane vacuum gravity as the set of
canonical coordinates and canonically conjugated momenta, 
\end{subequations}
\begin{subequations}
\label{constraintsV}
\begin{equation}
\{g_{A};\pi ^{A}\}:=\{g_{ab},M^{a},M;\pi ^{ab},p_{a},p\}\;,  \label{gApiA}
\end{equation}%
by introducing the notation $g_{A}=\{g_{ab},M^{a},M\}$ and $\pi ^{A}=\{\pi
^{ab},p_{a},p\}$ with the multi-index $A=1,2,3$ as a condensed notation for
the gravi-tensorial, gravi-vectorial and gravi-scalar degrees of freedom.
The momenta canonically conjugated to the field variables $g_{ab}$, $M^{a}$
and $M$ are 
\end{subequations}
\begin{subequations}
\label{momenta}
\begin{eqnarray}
\pi ^{ab}:= &&\frac{\partial \mathcal{L}^{G}}{\partial \dot{g}_{ab}}=M\sqrt{g%
}\left[ K^{ab}-\left( K+\mathcal{K}\right) g^{ab}\right] \ , \\
p_{a}:= &&\frac{\partial \mathcal{L}^{G}}{\partial \dot{M}^{a}}=2\sqrt{g}%
\mathcal{K}_{a}\ , \\
p:= &&\frac{\partial \mathcal{L}^{G}}{\partial \dot{M}}=-2\sqrt{g}K\ .
\end{eqnarray}%
Inverting Eqs. (\ref{momenta}) with respect to the second and normal
fundamental forms and the normal fundamental scalar, we obtain 
\end{subequations}
\begin{subequations}
\label{extrcurv}
\begin{eqnarray}
K_{ab} &=&\frac{1}{\sqrt{g}M}\left( \pi _{ab}-\frac{1}{s}g_{ab}\pi \right) -%
\frac{p}{2s\sqrt{g}}g_{ab}\ , \\
\mathcal{K}^{a} &=&\frac{p^{a}}{2\sqrt{g}}\ , \\
\mathcal{K} &=&\frac{s-1}{2s\sqrt{g}}p-\frac{\pi }{sM\sqrt{g}}\ .
\end{eqnarray}%
We insert these formulae into the Lagrangian (\ref{LG}), so that the
Einstein-Hilbert action takes the \char`\"{}already Hamiltonian form\char`\"{%
}: 
\end{subequations}
\begin{eqnarray}
&&S^{G}[g_{A},\pi ^{A};N^{a},N]=\int_{-\infty }^{\infty }dt\int_{-\infty
}^{\infty }d\chi \int_{\Sigma _{t\chi }}d^{s}x  \notag \\
&&\times (\pi ^{A}\dot{g}_{A}-N\mathcal{H}_{\bot }^{G}[g_{A},\pi ^{A}]-N^{a}%
\mathcal{H}_{a}^{G}[g_{A},\pi ^{A}])  \notag \\
&&-ST_{t=\pm \infty }-ST_{\partial \Sigma _{t\chi }}+ST_{\chi =\pm \infty }\
.  \label{actioncan}
\end{eqnarray}%
The extremization of the action (\ref{actioncan}) with respect to the lapse
and shift gives the constraint equations%
\begin{equation}
\mathcal{H}_{\bot }^{G}=-\frac{\delta S^{G}}{\delta N}\;,\quad \mathcal{H}%
_{a}^{G}=-\frac{\delta S^{G}}{\delta N^{a}}\;.  \label{constraintsfromaction}
\end{equation}%
In detail, the super-Hamiltonian constraint $\mathcal{H}^{G}$ and the
super-momentum constraint $\mathcal{H}_{a}^{G}$ of vacuum gravity, written
in terms of the canonical data, \ are: 
\begin{subequations}
\label{constrG}
\begin{eqnarray}
\mathcal{H}_{\bot }^{G}[g_{A},\pi ^{A}] &=&-\sqrt{g}[M(R-L^{2}+3L_{ab}L^{ab})
\notag \\
&&-2g^{ab}(\partial /\partial \chi -\pounds _{\mathbf{M}%
})L_{ab}-2D_{a}D^{a}M]  \notag \\
&&+\frac{1}{M\sqrt{g}}\left( \pi _{ab}\pi ^{ab}-\frac{1}{s}\pi ^{2}\right) -%
\frac{p\pi }{s\sqrt{g}}\   \notag \\
&&+\frac{M}{2\sqrt{g}}\left( p^{a}p_{a}+\frac{s-1}{2s}p^{2}\right) \ ,
\label{constrham} \\
\mathcal{H}_{a}^{G}[g_{A},\pi ^{A}] &=&-2D_{b}\pi ^{b}{}_{a}-(\partial
/\partial \chi -\pounds _{\mathbf{M}})p_{a}  \notag \\
&&+pD_{a}M~.  \label{constrdiffa}
\end{eqnarray}%
~ The contributions $ST_{t=\pm \infty }$, $ST_{\partial \Sigma _{t\chi }}$
and $ST_{\chi =\pm \infty }$ in the action (\ref{actioncan}) denote a
collection of surface terms on $t\rightarrow \pm \infty $, $\partial \Sigma
_{t\chi }$ and $\chi \rightarrow \pm \infty $ , respectively. Their
contribution can be compensated by adding surface terms to the action. The
contributions $\partial \Sigma _{t\chi }$ and $\chi \rightarrow \pm \infty $
do not have analogues in the standard ADM decomposition; and they come from
partial integrations meant to transform the set of variables $(N,~N^{a})$
into Lagrange-multipliers in the action. For completeness, we enlist these
terms: 
\end{subequations}
\begin{subequations}
\begin{eqnarray}
ST_{t=\pm \infty } &=&\frac{2}{s}\int_{-\infty }^{\infty }dt\int_{-\infty
}^{\infty }d\chi \int_{\Sigma _{t\chi }}d^{s}x  \notag \\
&&\times \frac{\partial }{\partial t}\left( \pi +\frac{M}{2}p\right) ~, \\
ST_{\partial \Sigma _{t\chi }} &=&2\int_{-\infty }^{\infty }dt\int_{-\infty
}^{\infty }d\chi \int_{\Sigma _{t\chi }}d^{s}xD_{a}\left[ N_{b}\pi
^{ab}\right.  \notag \\
&&-\frac{1}{s}N^{a}\left( \pi +\frac{M}{2}p\right) -\frac{1}{2}%
N^{b}p_{b}M^{a}  \notag \\
&&+\left. \sqrt{g}\left( N\mathcal{L}M^{a}+MD^{a}N\right) \right] ~, \\
ST_{\chi =\pm \infty } &=&2\int_{-\infty }^{\infty }\!dt\int_{-\infty
}^{\infty }\!d\chi \int_{\Sigma _{t\chi }}d^{s}x  \notag \\
&&\times \frac{\partial }{\partial \chi }\left( N\sqrt{g}\mathcal{L}-N^{a}%
\frac{p_{a}}{2}\right) ~.
\end{eqnarray}

The Hamiltonian of vacuum gravity can be written as the smearing of the
super-Hamiltonian and super-momentum constraints with the lapse function and
shift vector, 
\end{subequations}
\begin{equation}
H^{G}[N]=H_{\bot }^{G}[N]+H_{a}^{G}[N^{a}]\;,
\end{equation}%
where 
\begin{subequations}
\begin{eqnarray}
H_{\bot }^{G}[N] &=&\int d\chi \int_{\Sigma _{t\chi }}dx^{s}N(x,\chi )%
\mathcal{H}_{\bot }^{G}(x,\chi )\;, \\
H_{a}^{G}[N^{a}] &=&\int d\chi \int_{\Sigma _{t\chi }}dx^{s}N^{a}(x,\chi )%
\mathcal{H}_{a}^{G}(x,\chi )\;.
\end{eqnarray}%
The Poisson bracket of any two functions $f(x,\chi ;g_{A},\pi ^{A}]$ and $%
g(x,\chi ;g_{A},\pi ^{A}]$ on the phase space is defined as 
\end{subequations}
\begin{subequations}
\begin{gather}
\{f(x,\chi ),h(x^{\prime },\chi ^{\prime })\}=  \notag \\
\qquad \int d\chi ^{\prime \prime }\int_{\Sigma _{t\chi }}d^{s}x^{\prime
\prime }\frac{\delta f(x,\chi )}{\delta g_{A}(x^{\prime \prime },\chi
^{\prime \prime })}\frac{\delta h(x^{\prime },\chi ^{\prime })}{\delta \pi
^{A}(x^{\prime \prime },\chi ^{\prime \prime })}  \notag \\
\qquad -\int d\chi ^{\prime \prime }\int_{\Sigma _{t\chi }}d^{s}x^{\prime
\prime }\frac{\delta f(x,\chi )}{\delta \pi ^{A}(x^{\prime \prime },\chi
^{\prime \prime })}\frac{\delta h(x^{\prime },\chi ^{\prime })}{\delta
g_{A}(x^{\prime \prime },\chi ^{\prime \prime })}\;\;,\;
\end{gather}%
which provides the following non-vanishing commutation relations for the
Poisson brackets of the canonical variables: 
\end{subequations}
\begin{subequations}
\begin{eqnarray}
\{g_{ab}(x,\chi ),\pi ^{cd}(x^{\prime },\chi ^{\prime })\} &=&\delta
^{c}{}_{(a}\delta ^{d}{}_{b)}\delta (x,\chi ;x^{\prime },\chi ^{\prime
})\;\;, \\
\{M^{a}(x,\chi ),p_{b}(x^{\prime },\chi ^{\prime })\} &=&\delta
^{a}{}_{b}\delta (x,\chi ;x^{\prime },\chi ^{\prime })\;\;, \\
\{M(x,\chi ),p(x^{\prime },\chi ^{\prime })\} &=&\delta (x,\chi ;x^{\prime
},\chi ^{\prime })\;\;.
\end{eqnarray}%
or with the condensed notation 
\end{subequations}
\begin{equation}
\{g_{A}(x,\chi ),\pi ^{B}(x^{\prime },\chi ^{\prime })\}=\delta
^{B}{}_{A}\delta (x,\chi ;x^{\prime },\chi ^{\prime })\;\;.
\end{equation}%
Here $\delta ^{B}{}_{A}=\{\delta ^{c}{}_{(a}\delta ^{d}{}_{b)},\delta
^{a}{}_{b},1\}$, $\delta (x,\chi ;x^{\prime },\chi ^{\prime })=\delta
(x,x^{\prime })\delta (\chi ,\chi ^{\prime })$ and the Dirac delta
distribution behaves under coordinate transformations as a scalar in its
non-primed arguments, and as a scalar density of weight one in its primed
ones.

The constraints represent restrictions on the initial data and their Poisson
brackets \char`\"{}close\char`\"{} according to the Dirac algebra: 
\begin{subequations}
\label{Dirac}
\begin{align}
\{\mathcal{H}_{\bot }^{G}(x,\chi ),\mathcal{H}_{\bot }^{G}(x^{\prime },\chi
^{\prime })\}& =g^{ab}(x,\chi )\mathcal{H}_{a}^{G}(x,\chi )\delta
_{,b}(x,\chi ;x^{\prime },\chi ^{\prime })  \notag \\
& -(x\chi \leftrightarrow x^{\prime }\chi ^{\prime })~, \\
\{\mathcal{H}_{\bot }^{G}(x,\chi ),\mathcal{H}_{a}^{G}(x^{\prime },\chi
^{\prime })\}& =\mathcal{H}_{\bot }^{G}(x,\chi )\delta _{,a}(x,\chi
;x^{\prime },\chi ^{\prime })  \notag \\
& +\!\mathcal{H}_{,a}^{G}(x,\chi )\delta (x,\chi ;x^{\prime },\chi ^{\prime
})~, \\
\{\mathcal{H}_{a}^{G}(x,\chi ),\mathcal{H}_{b}^{G}(x^{\prime },\chi ^{\prime
})\}& =\mathcal{H}_{b}^{G}\delta _{,a}(x,\chi ;x^{\prime },\chi ^{\prime }) 
\notag \\
& -(ax\chi \leftrightarrow bx^{\prime }\chi ^{\prime })~.
\end{align}

Time evolution of the canonical data is generated by the Hamiltonian of the
system as 
\end{subequations}
\begin{subequations}
\label{canonical}
\begin{eqnarray}
\dot{g}_{A}(x,\chi )\! &=&\!\{g_{A}(x,\chi ),H^{G}[N]\}\!=\frac{\delta
H^{G}[N]}{\delta \pi ^{A}(x,\chi )}\;, \\
\dot{\pi}^{A}(x,\chi )\! &=&\!\{\pi ^{A}(x,\chi ),H^{G}[N]\}\!=-\frac{\delta
H^{G}[N]}{\delta g_{A}(x,\chi )}\;.
\end{eqnarray}%
Computation gives the dynamical equations for the gravi-tensor, gravi-vector
and gravi-scalar degrees of freedom: \textbf{\ } 
\end{subequations}
\begin{subequations}
\label{gdotsG}
\begin{eqnarray}
\dot{g}_{ab} &=&\frac{2N}{\sqrt{g}}\left[ \frac{1}{M}\left( \pi _{ab}-\frac{1%
}{s}\pi g_{ab}\right) -\frac{1}{2s}pg_{ab}\right]  \notag \\
&&+\pounds _{\mathbf{N}}g_{ab}\;\;,  \label{dotgab} \\
\dot{M}^{a} &=&\frac{MN}{\sqrt{g}}p^{a}+\frac{\partial N^{a}}{\partial \chi }%
+\pounds _{\mathbf{N}}M^{a}\;\;,  \label{dotMa} \\
\dot{M} &=&\frac{MN}{s\sqrt{g}}\left( \frac{s-1}{2}p-\frac{1}{M}\pi \right) +%
\pounds _{\mathbf{N}}M  \label{dotM}
\end{eqnarray}%
and 
\end{subequations}
\begin{subequations}
\label{pidotsG}
\begin{eqnarray}
\dot{\pi}^{ab} &=&N\mathcal{S}^{ab}+N\mathcal{V}^{ab}-NM\sqrt{g}\mathcal{L}%
(L^{ab}-Lg^{ab})  \notag \\
&&+\sqrt{g}(D^{a}D^{b}N-g^{ab}D^{c}D_{c}N-g^{ab}D_{c}ND^{c}M)  \notag \\
&&+\sqrt{g}g^{ab}(\partial /\partial \chi -\pounds _{\mathbf{M}})(N\mathcal{L%
})+\pounds _{\mathbf{N}}\pi ^{ab}\;\;,  \label{dotpiab} \\
\dot{p}_{a} &=&N\mathcal{V}_{a}-2\sqrt{g}[L^{b}{}_{a}D_{b}N+D_{a}(N\mathcal{L%
})]  \notag \\
&&+\pounds _{\mathbf{N}}p_{a}\;\;,  \label{dotpa} \\
\dot{p} &=&N\mathcal{S}+N\mathcal{V}-2\sqrt{g}(L\mathcal{L}+D_{a}D^{a}N) 
\notag \\
&&+\pounds _{\mathbf{N}}p~.  \label{dotp}
\end{eqnarray}%
Here $\mathcal{S}^{ab}$ and $\mathcal{S}$ denote the tensorial and scalar
projections\footnote{%
Note that no vectorial projection $\mathcal{S}_{a}$ is present in $\dot{p}%
_{a}$. However as time-derivatives and index raising do not commute, there
will be vectorial kinetic-type terms in $\dot{p}^{a}$.} of the geodesic
spray of the DeWitt super-metric \cite{FischerMarsden}:

\end{subequations}
\begin{subequations}
\label{S}
\begin{eqnarray}
\mathcal{S}^{ab}(\pi ^{A},\pi ^{A}) &=&-\frac{2}{M\sqrt{g}}\left( \pi
^{a}{}_{c}\pi ^{bc}-\frac{1}{s}\pi \pi ^{ab}\right)  \notag \\
&&+\frac{1}{2M\sqrt{g}}\left( \pi _{cd}\pi ^{cd}-\frac{1}{s}\pi ^{2}\right)
g^{ab}  \notag \\
&&-\frac{M}{2\sqrt{g}}g^{ab}\left( \frac{1}{sM}\pi p-\frac{1}{2}p_{c}p^{c}-%
\frac{s-1}{4s}p^{2}\right)  \notag \\
&&+\frac{1}{\sqrt{g}}\left( \frac{1}{s}p\pi ^{ab}+\frac{M}{2}%
p^{a}p^{b}\right) ~, \\
\mathcal{S}(\pi ^{A},\pi ^{A}) &=&\frac{1}{\sqrt{g}M^{2}}\left( \pi _{ab}\pi
^{ab}-\frac{1}{s}\pi ^{2}\right)  \notag \\
&&-\frac{1}{\sqrt{g}}\left( \frac{1}{2}p_{a}p^{a}+\frac{s-1}{4s}p^{2}\right)
~,
\end{eqnarray}%
\newline
while $\mathcal{V}^{ab},~\mathcal{V}_{a}$ and $\mathcal{V}$ represent the
tensorial, vectorial and scalar projections of the force term of the $\left(
s+1\right) $-dimensional scalar curvature potential: 
\end{subequations}
\begin{subequations}
\label{V}
\begin{eqnarray}
\mathcal{V}^{ab}(g_{A}) &=&-M\sqrt{g}\left(
G^{ab}+2L^{ac}L^{b}{}_{c}-LL^{ab}\right)  \notag \\
&&-\frac{1}{2}M\sqrt{g}\left( 3L^{cd}L_{cd}-L^{2}\right) g^{ab}  \notag \\
&&+\sqrt{g}\left( g^{ac}g^{bd}-g^{ab}g^{cd}\right) \left( \partial /\partial
\chi -\pounds _{\mathbf{N}}\right) L_{cd}  \notag \\
&&+\sqrt{g}(D^{a}D^{b}M-g^{ab}D^{c}D_{c}M)~, \\
\mathcal{V}_{a}(g_{A}) &=&-2\sqrt{g}(D_{b}L^{b}{}_{a}-D_{a}L)~, \\
\mathcal{V}(g_{A}) &=&\sqrt{g}(R+L_{ab}L^{ab}-L^{2})~.
\end{eqnarray}%
The evolution equations (\ref{pidotsG}) together with the constraints (\ref%
{constrG}) are equivalent with the vacuum Einstein equations in the bulk.
Once the constraints are obeyed at some instant of time, the dynamical
equations assure that they will continue to be satisfied later on.

\section{Hamiltonian dynamics with matter sources}

\subsection{General considerations}

The basic scheme of the Hamiltonian formulation does not change if we couple
matter fields to gravity. We only have to enlarge the phase space with the
canonical variables of the matter sources. The total action describing the
system is 
\end{subequations}
\begin{equation}
S=S^{G}\left[ \widetilde{g}_{ab}\right] +2\widetilde{\kappa }^{2}S^{M}\left[ 
\widetilde{g}_{ab},\Psi _{i}\right] ~  \label{totalaction}
\end{equation}%
where $\widetilde{\kappa }^{2}$ is the gravitational coupling constant in $%
s+2$ dimensions and we assume the matter action $S^{M}\left[ \widetilde{g}%
_{ab},\Psi _{i}\right] $ contains the metric only in non-derivative terms.
This assumption is obeyed for all physically relevant matter fields and it
assures that (a) the vacuum gravitational momenta (\ref{momenta}) remain
unchanged in the presence of matter; (b) by performing the Legendre
transformation, the matter contribution to the total Hamiltonian is just
minus its contribution to the total Lagrangian.

Extremizing the matter action with respect to the matter fields $\Psi _{i}$
yields their evolution equations. Extremizing with respect to the metric
yields, by definition, the energy-momentum tensor%
\begin{equation}
\widetilde{T}^{ab}=\frac{2}{\sqrt{-\widetilde{g}}}\frac{\delta S^{M}}{\delta 
\widetilde{g}_{ab}}~.
\end{equation}

We can perform the $\left( s+1+1\right) $-decomposition of the matter
Lagrangian density, without specifying it explicitly. For this we write the
energy-momentum tensor as%
\begin{align}
\widetilde{T}^{ab}& =\left( \widetilde{T}^{cd}g_{c}^{a}g_{d}^{b}\right)
+\left( \widetilde{T}^{cd}n_{c}n_{d}\right) n^{a}n^{b}  \notag \\
& +\left( \widetilde{T}^{cd}l_{c}l_{d}\right) l^{a}l^{b}-2\left( \widetilde{T%
}^{cd}g_{c}^{(a}n_{d}\right) n^{b)}  \notag \\
& +2\left( \widetilde{T}^{cd}g_{c}^{(a}l_{d}\right) l^{b)}-2\left( 
\widetilde{T}^{cd}n_{c}l_{d}\right) n^{(a}l^{b)}~,
\end{align}%
and we note that the decomposition Eq. (\ref{tildeg0}) of the metric gives%
\begin{equation}
\delta \widetilde{g}_{ab}=\delta g_{ab}-2n_{(a}\delta n_{b)}+2l_{(a}\delta
l_{b)}~.
\end{equation}%
The variation of the matter action with respect to the metric (after partial
integrations) results in 
\begin{gather}
\delta _{\widetilde{g}}S^{M}=\int d^{s+2}x~\frac{\delta S^{M}}{\delta 
\widetilde{g}_{ab}}\delta \widetilde{g}_{ab}=  \notag \\
\int d^{s+2}xNM\sqrt{g}\left\{ \frac{1}{2}\left( \widetilde{T}%
^{cd}g_{c}^{a}g_{d}^{b}\right) \delta g_{ab}\right.  \notag \\
\quad -\left[ \widetilde{T}^{cd}g_{c}^{a}n_{d}-\left( \widetilde{T}%
^{cd}n_{c}n_{d}\right) n^{a}+\left( \widetilde{T}^{cd}n_{c}l_{d}\right) l^{a}%
\right] \delta n_{a}  \notag \\
\quad +\left[ \widetilde{T}^{cd}g_{c}^{a}l_{d}-\left( \widetilde{T}%
^{cd}n_{c}l_{d}\right) n^{a}+\left( \widetilde{T}^{cd}l_{c}l_{d}\right) l^{a}%
\right] \delta l_{a}  \notag \\
\quad \left. +\left( \widetilde{T}^{cd}g_{bc}n_{d}\right) \delta
n^{b}-\left( \widetilde{T}^{cd}g_{bc}l_{d}\right) \delta l^{b}\right\} \;.
\end{gather}%
We would like to replace the variations $\delta n^{a},~\delta l^{a},~\delta
n_{a},~\delta l_{a}$ by the variation of our chosen gravitational variables.
As $\partial /\partial t$ and $\partial /\partial \chi $ are directions
unaffected by the variation of the metric, from Eqs. (\ref{off-brane}) and (%
\ref{temporal}) we obtain 
\begin{subequations}
\begin{eqnarray}
\delta n^{a} &=&-\left( \frac{\delta N}{N}n^{a}+\frac{\delta N^{a}}{N}%
\right) \ , \\
\delta l^{a} &=&-\left( \frac{\delta M}{M}l^{a}+\frac{\delta M^{a}}{M}%
\right) \ .
\end{eqnarray}%
As, cf. to Paper \textbf{I} , the dual bases are related as $\left(
dt\right) _{a}=n_{a}/N$ and $\left( d\chi \right) _{a}=l_{a}/M$, the
variation of the co-vectors also arises: 
\end{subequations}
\begin{equation}
\delta n_{a}=\frac{\delta N}{N}n_{a}~,\text{ \ \ \ \ \ \ \ \ \ \ \ }\delta
l_{a}=\frac{\delta M}{M}l_{a}~,
\end{equation}%
and we obtain the desired formula:%
\begin{eqnarray}
&&\delta _{\widetilde{g}}S^{M}[g_{ab},M^{a},M;N^{a},N;~\Psi ]=  \notag \\
&&\quad \int dt\int d\chi \int_{\Sigma _{t\chi }}d^{s}x\sqrt{g}\left\{ \frac{%
NM}{2}\left( \widetilde{T}^{cd}g_{c}^{a}g_{d}^{b}\right) \delta g_{ab}\right.
\notag \\
&&\quad +N\left( \widetilde{T}^{cd}l_{c}l_{d}\right) \delta M+N\left( 
\widetilde{T}^{cd}g_{c}^{b}l_{d}\right) g_{ab}\delta M^{a}  \notag \\
&&\quad \left. -M\left( \widetilde{T}^{cd}n_{c}n_{d}\right) \delta N-M\left( 
\widetilde{T}^{cd}g_{c}^{b}n_{d}\right) g_{ab}\delta N^{a}\right\} ~,
\label{actionMatterCan}
\end{eqnarray}%
(Due to the non-derivative coupling of matter to gravity there are no
momenta dependencies.) The result (\ref{actionMatterCan}) shows that
extremizing the total action with respect to the lapse function $N$ and the
shift vector $N^{a}$ (similarly to the prescription (\ref%
{constraintsfromaction})) gives the super-Hamiltonian and super-momentum
contribution of the matter fields 
\begin{subequations}
\label{constrM}
\begin{eqnarray}
\mathcal{H}_{\bot }^{M} &=&2\widetilde{\kappa }^{2}M\sqrt{g}\left( 
\widetilde{T}^{cd}n_{c}n_{d}\right) ~,  \label{constrhamM} \\
\mathcal{H}_{a}^{M} &=&2\widetilde{\kappa }^{2}M\sqrt{g}\left( \widetilde{T}%
^{cd}g_{c}^{b}n_{d}\right) g_{ab}\!~.  \label{constrdiffaM}
\end{eqnarray}%
The super-Hamiltonian and super-momentum constraints of the total system can
be written as 
\end{subequations}
\begin{subequations}
\label{constrGM}
\begin{eqnarray}
\mathcal{H}_{\bot } &=&\mathcal{H}_{\bot }^{G}+\mathcal{H}_{\bot
}^{M}\approx 0~,  \label{totalHamconstr} \\
\mathcal{H}_{a} &=&\mathcal{H}_{a}^{G}+\mathcal{H}_{a}^{M}\approx 0~,
\label{totaldiffconstr}
\end{eqnarray}%
with the vacuum and matter contributions given by Eqs. (\ref{constrG}) and
Eqs. (\ref{constrM}). Here $\approx $ denotes weak equality (holding on the
constraint surface in the phase space).

In what follows, we discuss the canonical equations in the presence of
matter. Due to the non-derivative coupling, Eqs. (\ref{gdotsG}) remain valid
in the presence of matter (since $\delta S^{M}/\delta \pi ^{A}=0$). However
the evolution of the momenta receive additional contributions. Due to remark
(b), the matter contributions to the left hand side of Eqs. (\ref{pidotsG})
can be found by extremizing the action with respect to the dynamical
variables $g_{ab},~M^{a}$ and $M$: 
\end{subequations}
\begin{subequations}
\label{dSM}
\begin{eqnarray}
\frac{\delta S^{M}}{\delta g_{ab}} &=&\frac{NM}{2}\sqrt{g}\left( \widetilde{T%
}^{cd}g_{c}^{a}g_{d}^{b}\right) \;,  \label{dSMdgab} \\
\frac{\delta S^{M}}{\delta M^{a}} &=&N\sqrt{g}\left( \widetilde{T}%
^{cd}g_{ac}l_{d}\right) \;,  \label{dSMdMa} \\
\frac{\delta S^{M}}{\delta M} &=&N\sqrt{g}\left( \widetilde{T}%
^{cd}l_{c}l_{d}\right) \;.  \label{dSMdM}
\end{eqnarray}%
The dynamical equations for $\pi ^{A}$ with the contributions (\ref{dSM})
take the form 
\end{subequations}
\begin{subequations}
\label{dotpiGMgen}
\begin{eqnarray}
\dot{\pi}^{ab} &=&N\mathcal{S}^{ab}+N\mathcal{V}^{ab}-NM\sqrt{g}\mathcal{L}%
(L^{ab}-Lg^{ab})  \notag \\
&&+\sqrt{g}(D^{a}D^{b}N-g^{ab}D^{c}D_{c}N-g^{ab}D_{c}ND^{c}M)  \notag \\
&&+\sqrt{g}g^{ab}(\partial /\partial \chi -\pounds _{\mathbf{M}})(N\mathcal{L%
})+\pounds _{\mathbf{N}}\pi ^{ab}  \notag \\
&&+\widetilde{\kappa }^{2}NM\sqrt{g}\left( \widetilde{T}%
^{cd}g_{c}^{a}g_{d}^{b}\right) \;\;,  \label{dotpiab1} \\
\dot{p}_{a} &=&N\mathcal{V}_{a}N\mathcal{V}_{a}-2\sqrt{g}%
[L^{b}{}_{a}D_{b}N+D_{a}(N\mathcal{L})]  \notag \\
&&+\pounds _{\mathbf{N}}p_{a}+2\widetilde{\kappa }^{2}N\sqrt{g}\left( 
\widetilde{T}^{bc}g_{ab}l_{c}\right) \;\;,  \label{dotpa1} \\
\dot{p} &=&N\mathcal{S}+N\mathcal{V}-2\sqrt{g}(L\mathcal{L}+D_{a}D^{a}N) 
\notag \\
&&+\pounds _{\mathbf{N}}p+2\widetilde{\kappa }^{2}N\sqrt{g}\left( \widetilde{%
T}^{ab}l_{a}l_{b}\right) \;\;.  \label{dotp1}
\end{eqnarray}%
These formulae are valid for any matter source coupled non-derivatively to
the $\left( s+2\right) $-geometry. The matter contributions can be further
specified, once the energy-momentum tensor (or equivalently, the matter
Lagrangian) is known.

\subsection{Brane-world scenario}

In the brane-world scenarios the stress-energy tensor is decomposed as 
\end{subequations}
\begin{equation}
\widetilde{T}_{ab}=\widetilde{\Pi }_{ab}+[-\lambda
\,{}^{(s+1)}g_{ab}+{}^{\left( s+1\right) }T_{ab}]\delta (\chi )\;,
\label{tildeTab}
\end{equation}%
where the regular part $\widetilde{\Pi }_{ab}$ represents the non-standard
model bulk sources, while the distributional term contains the brane tension 
$\lambda $ and the energy-momentum tensor of standard model matter field
localized on the brane. First we decompose the bulk energy momentum $%
\widetilde{\Pi }_{ab}$ with respect to the off-brane normal $l^{a}$ as: 
\begin{equation}
\frac{s-1}{s}\widetilde{\Pi }_{ab}=~^{\left( s+1\right) }\mathcal{P}%
_{ab}+2l_{(a}~^{\left( s+1\right) }\mathfrak{P}_{b)}+l_{a}l_{b}\mathfrak{P~.}
\label{PIab}
\end{equation}%
Then we decompose $^{\left( s+1\right) }\mathcal{P}_{ab}$ and $^{\left(
s+1\right) }\mathfrak{P}_{b}$ further with respect to the time-like normal $%
n^{a}$ as: 
\begin{subequations}
\begin{eqnarray}
^{\left( s+1\right) }\mathcal{P}_{ab} &=&\mathcal{P}_{ab}+2n_{(a~}\mathcal{P}%
_{b)}+n_{a}n_{b}\mathcal{P},  \label{Pab} \\
^{\left( s+1\right) }\mathfrak{P}_{a} &=&\mathrm{P}_{a}+\mathrm{P}n_{a}~,
\label{Pb}
\end{eqnarray}%
where $\mathcal{P}_{ab}n^{b}=\mathcal{P}_{a}n^{a}=0=\mathrm{P}_{a}n^{a}$.
(Note that $g^{ab}\mathcal{P}_{ab}\neq \mathcal{P}$, unless $^{\left(
s+1\right) }\mathcal{P}_{ab}$ happens to be traceless.)

The brane contribution can be algebraically decomposed as 
\end{subequations}
\begin{equation}
^{\left( s+1\right) }T_{ab}=\rho n_{a}n_{b}+Pg_{ab}+\Pi _{ab}+2n_{(a}Q_{b)}
\label{Tab}
\end{equation}%
with respect to the 4-velocity $n^{a}$ of the fluid, and in terms of the
energy density $\rho $, isotropic pressure $P$, anisotropic stresses $\Pi
_{ab}$ and the energy transport (heat flow) $Q_{a}$ (here $g^{ab}\Pi
_{ab}=n^{a}\Pi _{ab}=n^{a}Q_{a}=0$).

Choosing the 4-velocity of the fluid as normal to the spatial slices does
not restrict the arbitrareness of the foliation. Indeed, the foliation is
given by the form $\left( dt\right) _{a}=n_{a}/N$, while the normal vector
field is $n^{a}=\widetilde{g}^{ab}\left( dt\right) _{b}/N$, which involves
the $\left( s+2\right) $-metric, and as such, the lapse and shift. The
arbitrareness of the lapse function and shift vector assures that one can
still choose various foliations once $n^{a}$ is fixed by the chosen
reference fluid. The time parameter $t_{ref}$ defined by the fluid as $%
n^{a}=(\partial /\partial t_{ref})^{a}$ is different from the time $t\ $%
defined by the chosen foliation. Therefore while we associate $t_{ref}$ with
the cosmological time, we still have the freedom of evolving the system with
respect to any conveniently chosen time parameter $t$ (like the conformal
time). Restricting $N=1$ and $N^{a}=0$ leads to the identification $%
t=t_{ref.}$.

The above conditions with Eqs. (\ref{constrM}) and Eqs. (\ref{dSM}) give 
\begin{subequations}
\label{constrM2}
\begin{eqnarray}
&&\mathcal{H}_{\bot }^{M}[g_{A};\mathcal{P},\rho ]=  \notag \\
&&\qquad 2\widetilde{\kappa }^{2}M\sqrt{g}\left[ \frac{s}{s-1}\mathcal{P}%
+(\rho +\lambda )\delta (\chi )\right] ~,  \label{constrhamM2} \\
&&\mathcal{H}_{a}^{M}[g_{A};\mathcal{P}_{a},Q_{a}]=  \notag \\
&&\qquad -2\widetilde{\kappa }^{2}M\sqrt{g}\left[ \frac{s}{s-1}\mathcal{P}%
_{a}+Q_{a}\delta (\chi )\right] ~,  \label{constrdiffM2}
\end{eqnarray}%
and 
\end{subequations}
\begin{subequations}
\label{dSM2}
\begin{align}
\frac{\delta S^{M}}{\delta g_{ab}}& =\frac{NM}{2}\sqrt{g}\frac{s}{s-1}%
\mathcal{P}^{ab}  \notag \\
& +\frac{NM}{2}\sqrt{g}\left[ \Pi ^{ab}+\left( P-\lambda \right) \,{}g^{ab}%
\right] \delta (\chi )\;,\quad  \label{dSMdg2} \\
\frac{\delta S^{M}}{\delta M^{a}}& =N\sqrt{g}\frac{s}{s-1}\mathrm{P}_{a}\;,
\\
\frac{\delta S^{M}}{\delta M}& =N\sqrt{g}\frac{s}{s-1}\mathfrak{P}\;.
\label{dSM2M2}
\end{align}

The full brane-world geometrodynamics in the presence of matter is then
given by the equations (\ref{gdotsG}) for $\dot{g}_{A}$ and the sum of the
right hand sides of Eqs. (\ref{pidotsG}) and (\ref{dSM2}) for $\dot{\pi}^{A}$%
: 
\end{subequations}
\begin{subequations}
\label{pidotsGM}
\begin{eqnarray}
\dot{\pi}^{ab} &=&N\mathcal{S}^{ab}+N\mathcal{V}^{ab}-NM\sqrt{g}\mathcal{L}%
(L^{ab}-Lg^{ab})  \notag \\
&&+\sqrt{g}(D^{a}D^{b}N-g^{ab}D^{c}D_{c}N-g^{ab}D_{c}ND^{c}M)  \notag \\
&&+\sqrt{g}g^{ab}(\partial /\partial \chi -\pounds _{\mathbf{M}})(N\mathcal{L%
})+\pounds _{\mathbf{N}}\pi ^{ab}  \notag \\
&&+\widetilde{\kappa }^{2}NM\sqrt{g}\left\{ \frac{s}{s-1}\mathcal{P}%
^{ab}\right.  \notag \\
&&\left. +\left[ \Pi ^{ab}+\left( P-\lambda \right) \,{}g^{ab}\right] \delta
(\chi )\right\} \;,  \label{dotpiab2} \\
\dot{p}_{a} &=&N\mathcal{V}_{a}-2\sqrt{g}[L^{b}{}_{a}D_{b}N+D_{a}(N\mathcal{L%
})]  \notag \\
&&+\pounds _{\mathbf{N}}p_{a}+\widetilde{\kappa }^{2}N\sqrt{g}\frac{2s}{s-1}%
\mathrm{P}_{a}\;,  \label{dotpa2} \\
\dot{p} &=&N\mathcal{S}+N\mathcal{V}-2\sqrt{g}(L\mathcal{L}+D_{a}D^{a}N) 
\notag \\
&&+\pounds _{\mathbf{N}}p+\widetilde{\kappa }^{2}N\sqrt{g}\frac{2s}{s-1}%
\mathfrak{P}~.  \label{dotp2}
\end{eqnarray}

The constraints (\ref{totalHamconstr}) and (\ref{totaldiffconstr}) take the
form 
\end{subequations}
\begin{subequations}
\label{totalconstr}
\begin{eqnarray}
&&0\approx \mathcal{H}_{\bot }[g_{A},\pi ^{A};\rho ,\mathcal{P}]=  \notag \\
&&\quad -\sqrt{g}M(R-L^{2}+3L_{ab}L^{ab})  \notag \\
&&\quad -2\sqrt{g}\left[ g^{ab}(\partial /\partial \chi -\pounds _{\mathbf{M}%
})L_{ab}-D_{a}D^{a}M\right]  \notag \\
&&\quad +\frac{1}{M\sqrt{g}}\left( \pi _{ab}\pi ^{ab}-\frac{1}{s}\pi
^{2}\right) -\frac{p\pi }{s\sqrt{g}}  \notag \\
&&\quad +\frac{M}{2\sqrt{g}}\left( p^{a}p_{a}+\frac{s-1}{2s}p^{2}\right) \  
\notag \\
&&\ \quad +2\widetilde{\kappa }^{2}M\sqrt{g}\left[ \frac{s}{s-1}\mathcal{P}%
+(\rho +\lambda )\delta (\chi )\right] ~,  \label{totalHamconstr2} \\
&&0\approx \mathcal{H}_{a}[g_{A},\pi ^{A};\mathcal{P}_{a},Q_{a}]=  \notag \\
&&\quad -2D_{b}\pi ^{b}{}_{a}-(\partial /\partial \chi -\pounds _{\mathbf{M}%
})p_{a}+pD_{a}M  \notag \\
&&\quad -2\widetilde{\kappa }^{2}M\sqrt{g}\left[ \frac{s}{s-1}\mathcal{P}%
_{a}+Q_{a}\delta (\chi )\right] ~.  \label{totaldiffconstr2}
\end{eqnarray}%
The dynamical equations (\ref{gdotsG}) and (\ref{pidotsGM}) and the
constraints (\ref{totalconstr}) completely determine the time evolution of
the geometry and the matter fields on the brane in brane-world scenarios.
The LSDI matching condition follows from these equations, as we will show it
in the next section.

The $\delta $-function type distributional sources in the evolution
equations (\ref{dotpiab2}) need some further interpretation. Such
contributions also appear in the dynamics of $K_{ab}$ and $\mathcal{K}$, as
derived in Paper \textbf{I}. These contributions indicate the singular
behavior of $\dot{\pi}^{ab}$. This is however, not surprising. The canonical
equations and constraints are equivalent with the $\left( s+2\right) $%
-dimensional Einstein equations. If the sources of the latter are singular
(in the present case across the brane), the Riemann (and Einstein) tensors
are also singular, and certain singularities will be carried over in the
canonical equations. Traditionally (for example in the derivation of the
effective Einstein equation \cite{SMS}) the coefficients of the $\delta $%
-functions are interpreted as contributions present on the brane, but not in
the bulk regions.

\section{Regularization across the brane}

The brane contains $\delta $-function type distributional sources, which in
turn appear in both constraints (\ref{totalconstr}) and in the dynamical
equation (\ref{dotpiab2}). For these equations we apply the following
regularization procedure. First we consider a domain of finite thickness $%
\left( -\chi _{0},~\chi _{0}\right) $ enclosing the brane at $\chi =0$ and
we integrate the above-derived equations across its width. As a consequence,
the Dirac distribution $\delta \left( \chi \right) $ disappears. More
precisely, according to \cite{Decomp}, for any $\mathcal{H}\left( l\right) $
the relation 
\end{subequations}
\begin{equation}
\int_{-l_{0}}^{l_{0}}dl\delta \left( l\right) \mathcal{H}\left( l\right) =%
\mathcal{H}\left( 0\right)
\end{equation}%
holds. The integration across the brane is carried out here over a normal
coordinate $l$, defined as $\partial /\partial l=\mathbf{l}$ (see \cite%
{Decomp}). Should we employ the coordinate $\chi $, defined as $\partial
/\partial \chi =M\mathbf{l}+\mathbf{M}$ for integration across the brane
(thus $dl/d\chi =M$), we obtain 
\begin{equation}
\int_{-\chi _{0}}^{\chi _{0}}d\chi \delta \left( \chi \right) \mathcal{H}%
\left( \chi \right) =\frac{\mathcal{H}\left( 0\right) }{M}~.
\end{equation}%
Secondly, the primitive function of the integral of any total $\chi $%
-derivative term evaluated at the left and right domain boundaries $-\chi
_{0}$ and $\chi _{0}$ give the so-called jump of the respective quantities.
Thus, any quantity $\mathcal{G}$ appearing as $\partial \mathcal{G}/\partial
\chi $ in the respective equation leads to its jump across the brane $\Delta 
\mathcal{G}$, when the integration is carried out. Finally, if the value of $%
\chi _{0}$ is small, any other smooth function of $\chi _{0}$ can be
regarded as a constant, such that its integral will be proportional to the
width $2\chi _{0}$. When we take the thin brane limit $\chi _{0}\rightarrow
0 $, this procedure drops all such terms, and what remains are only the
terms originally multiplying $\delta $-functions and the jumps arising from
the total $\chi $-derivatives. If no such terms are present in any of the
equations derived in the preceding sections, we obtain identities (of $0=0$
type). Therefore non-trivial information arises only from the equations with
total $\chi $-derivatives and/or $\delta $-functions.

We can regain the junction condition for the embedding of the brane by
integrating the constraints and those dynamical equations which contain
derivatives with respect to $\chi $, the coordinate running in the off-brane
direction.

As Eqs. (\ref{constrham}) and (\ref{constrdiffa}) show, the vacuum
constraints contain the $\chi $-derivatives of $L_{ab}$ and $p_{a}$. The
integration of the super-Hamiltonian constraint (\ref{totalHamconstr})
provides 
\begin{equation}
\sqrt{g}g^{ab}\Delta L_{ab}=-\widetilde{\kappa }^{2}\lim_{\chi
_{0}\rightarrow 0}\int_{-\chi _{0}}^{\chi _{0}}d\chi M\sqrt{g}\left( 
\widetilde{T}^{cd}n_{c}n_{d}\right) \;.
\end{equation}%
By considering the brane world scenario, the integration of Eq. (\ref%
{totalHamconstr2}) simply gives the trace of the junction condition (53) of
Paper \textbf{I}: 
\begin{equation}
\Delta L=-\widetilde{\kappa }^{2}(\rho +\lambda )\;,  \label{DelL}
\end{equation}%
since $\int_{-\chi _{0}}^{\chi _{0}}d\chi f(\chi )\delta (\chi )=f(0)/M$.

When we integrate the supermomentum constraint (\ref{totaldiffconstr}), we
obtain%
\begin{equation}
\Delta p_{a}=2\widetilde{\kappa }^{2}\lim_{\chi _{0}\rightarrow
0}\int_{-\chi _{0}}^{\chi _{0}}d\chi M\sqrt{g}\left( \widetilde{T}%
^{cd}g_{c}^{b}n_{d}\right) g_{ab}\;,  \label{delpa}
\end{equation}%
since the integral of the finite terms in the momentum constraint vanishes
as $\chi _{0}\rightarrow 0$. For the matter fields on the brane specified in
Eq. (\ref{constrdiffM2}), the integration of the momentum constraint (\ref%
{totaldiffconstr2}) gives%
\begin{equation}
\Delta p_{a}=-2\widetilde{\kappa }^{2}\sqrt{g}Q_{a}\;,  \label{delpa2}
\end{equation}%
which is the vectorial projection (\ref{Lanczos_vector}) of the Lanczos
equation, rewritten in terms of momenta.

By integrating the dynamical equation (\ref{dotpiab1}) over $\chi $, we
obtain%
\begin{eqnarray}
&&\sqrt{g}\left[ \Delta L^{ab}-\Delta (L-\mathcal{L})g^{ab}\right]  \notag \\
&&\quad =-\widetilde{\kappa }^{2}\lim_{\chi _{0}\rightarrow 0}\int_{-\chi
_{0}}^{\chi _{0}}d\chi M\sqrt{g}\widetilde{T}^{cd}g_{c}^{a}g_{d}^{b}\;.
\end{eqnarray}%
whereas the integration of the dynamical equation (\ref{dotpiab2}) leads to
the expression 
\begin{equation}
\Delta L_{ab}-\Delta (L-\mathcal{L})g_{ab}=-\widetilde{\kappa }^{2}\left[
\Pi _{ab}+(P-\lambda )g_{ab}\right] \;.  \label{DelLab}
\end{equation}%
After inserting Eq. (\ref{DelL}) in the trace of this result, we get%
\begin{equation}
\Delta \mathcal{L}=\widetilde{\kappa }^{2}\frac{(1-s)\rho -sP+\lambda }{s}\;,
\label{delcalL}
\end{equation}%
which is the scalar projection (\ref{Lanczos_scalar}) of the Lanczos
equation.

The substitution of Eq. (\ref{delcalL}) into Eq. (\ref{DelLab}) gives%
\begin{equation}
\Delta L_{ab}=-\widetilde{\kappa }^{2}\left( \Pi _{ab}+\frac{\rho +\lambda }{%
s}g_{ab}\right) ~,  \label{delLab}
\end{equation}%
Then we have obtained the tensorial projection (\ref{Lanczos_tensor}) of the
Lanczos equation. This means the dynamical system (\ref{pidotsGM}) with the
constraints (\ref{totalconstr}) imply the usual LSDI junction conditions for
the brane.

By imposing $Z_{2}$ symmetry in the bulk across the brane (which implies $%
\Delta L_{ab}=2L_{ab}$, $\Delta p_{a}=2p_{a}$, and $\Delta \mathcal{L}=2%
\mathcal{L}$), we can express the components of the extrinsic curvature
associated with the brane normal in terms of the matter field variables: 
\begin{subequations}
\label{LanczosZ2}
\begin{eqnarray}
L_{ab} &=&-\frac{\widetilde{\kappa }^{2}}{2}\left( \Pi _{ab}+\frac{\rho
+\lambda }{s}g_{ab}\right) ~,  \label{Lab2} \\
p_{a} &=&-\widetilde{\kappa }^{2}\sqrt{g}Q_{a}\;,  \label{pa2} \\
\mathcal{L} &=&\widetilde{\kappa }^{2}\frac{(1-s)\rho -sP+\lambda }{2s}~.
\label{calL2}
\end{eqnarray}

As a simple application, we give the dynamical equation of the heat flow.
Eq. (\ref{pa2}) implies 
\end{subequations}
\begin{equation}
\widetilde{\kappa }^{2}\sqrt{g}\dot{Q}_{a}=\frac{p_{a}}{2}\dot{g}^{bc}\dot{g}%
_{bc}-\dot{p}_{a}\;.
\end{equation}%
Here both $\dot{g}_{bc}$ and $\dot{p}_{a}$ are known as Eqs. (\ref{dotgab})
and (\ref{dotpa2}). We obtain:%
\begin{eqnarray}
\widetilde{\kappa }^{2}\sqrt{g}\dot{Q}_{a} &=&p_{a}\left( -\frac{N}{2\sqrt{g}%
}p+D_{b}N^{b}\right)  \notag \\
&&-N\mathcal{V}_{a}+2\sqrt{g}[L^{b}{}_{a}D_{b}N+D_{a}(N\mathcal{L})]  \notag
\\
&&-\pounds _{\mathbf{N}}p_{a}-\widetilde{\kappa }^{2}N\sqrt{g}\frac{2s}{s-1}%
\mathrm{P}_{a}\;.
\end{eqnarray}%
This is the equation of heat flow expressed in terms of canonical data.

\section{Concluding remarks}

We have derived the Hamiltonian dynamics of the $\left( s+2\right) $%
-dimensional gravitation in terms of variables adapted to the existence of
the preferred \thinspace $s$-dimensional hypersurface. The canonical
(gravi-tensorial, gravi-vectorial and gravi-scalar) metric variables $%
g_{ab\,},~M^{a}$ and $M$ have canonically conjugated momenta $\pi
^{ab},~p_{a}$ and $p$, related to the extrinsic curvatures associated to the
temporal normal $n^{a}$. We have given the evolution equations for the
canonical data and also the super-Hamiltonian and super-momentum
constraints, all derived from an action principle.

Some of these equations contain $\delta $-function type contributions, due
to the singular source terms on the brane. These terms can be dropped, when
we monitor gravitational dynamics in the bulk, and kept on the brane.

The regularization of these equations across the brane yields the
projections of the Lanczos equation, written in terms of canonical data.

In the original covariant formulation of brane-world dynamics \cite{SMS} the
effective Einstein equation is obtained by expressing the terms quadratic in
the extrinsic curvatures with matter variables. This is achieved by
employing the Lanczos equation. In the present formalism, the role of these
extrinsic curvatures are taken by $L_{ab}$, $\mathcal{L}_{a}=\mathcal{K}_{a}$
and $\mathcal{L}$, all functions of the canonical variables. The projections
of the Lanczos equation derived in this paper can also be employed to
eliminate these geometrical variables in terms of matter variables in the
canonical equations.

As a simple application we have derived the equation of heat flow in terms
of canonical data.

The importance of the presented formalism relies in its possible application
in the initial-value problem in brane-worlds and in the prospect of
canonical quantization of brane-world gravity.

\section{Acknowledgments}

This work was supported by OTKA grants no. 46939 and 69036, the J\'{a}nos
Bolyai Fellowships of the Hungarian Academy of Sciences, the Pierre Auger
grant 05 CU 5PD1/2 via DESY/BMF and the EU Erasmus Collaboration between the
University of Szeged and the University of Bonn.

\appendix

\section{The $s+1+1$-Decomposition of energy-momentum and the Einstein
tensors}

In Appendix D of Paper \textbf{I} we have given the complete set of the
decompositions of the Riemann-, Ricci- and Einstein tensors. Due to a typo,
from the decomposition of the curvature scalar, Eq. (D3) of Paper \textbf{I}
the term $2\left( \mathcal{K}^{2}-\mathcal{L}^{2}\right) $ was omitted,
which cancels out the corresponding terms in the projections $%
g_{a}^{c}g_{b}^{d}\widetilde{G}_{cd}$, $n^{a}n^{b}\widetilde{G}_{ab}$ and $%
l^{a}l^{b}\widetilde{G}_{ab}$. Therefore the formulae (D3)-(D4c) of Paper 
\textbf{I} correctly read: 
\begin{widetext}
\begin{eqnarray}
\widetilde{R} & = & \widetilde{R}=R-3K_{ab}K^{ab}+K^{2}+2%
\left[(K+\mathcal{K})\mathcal{K}+\mathcal{K}_{a}\mathcal{K}%
^{a}+g^{ab}\pounds_{\mathbf{n}}K_{ab}+\pounds_{\mathbf{n}}\mathcal{K}\right]%
\notag\\
 &  & +3L_{ab}L^{ab}-L^{2}+2\left[(L-\mathcal{L})\mathcal{L}%
-g^{ab}\pounds_{\mathbf{l}}L_{ab}+\pounds_{\mathbf{l}}\mathcal{L}\right]%
\notag\\
 &  & -2\left[N^{-1}D_{a}D^{a}N+M^{-1}D_{a}D^{a}M+\left(NM%
\right)^{-1}D_{a}ND^{a}M\right]\label{tildeR}
\end{eqnarray}
and 
\begin{subequations}
\begin{eqnarray}
g_{a}^{c}g_{b}^{d}\widetilde{G}_{cd} & = &
G_{ab}-2K_{ac}K_{b}^{c}+\left(K+\mathcal{K}\right)K_{ab}-2\mathcal{K}_{a}%
\mathcal{K}_{b}+\pounds_{\mathbf{n}}K_{ab}-N^{-1}D_{b}D_{a}N\notag\\
 &  & +%
\left[\frac{1}{2}(3K_{cd}K^{cd}-K^{2})-(K+\mathcal{K})\mathcal{K}-\mathcal{K}%
_{c}\mathcal{K}^{c}-g^{cd}\pounds_{\mathbf{n}}K_{cd}-\pounds_{\mathbf{n}}%
\mathcal{K}+N^{-1}D_{c}D^{c}N\right]g_{ab}\notag\\
 &  &
+2L_{ac}L_{b}^{c}-\left(L-\mathcal{L}\right)L_{ab}-\pounds_{\mathbf{l}%
}L_{ab}-M^{-1}D_{b}D_{a}M\notag\\
 &  & -\left[\frac{1}{2}%
(3L_{cd}L^{cd}-L^{2})+\mathcal{L}(L-\mathcal{L})-g^{cd}\pounds_{\mathbf{l}%
}L_{cd}-\pounds_{\mathbf{l}}\mathcal{L}-M^{-1}D_{c}D^{c}M\right]%
g_{ab}\notag\\
 &  & +\left(NM\right)^{-1}D_{c}ND^{c}M\;\;,\label{ggG}%
\\
g_{a}^{c}n^{d}\widetilde{G}_{cd} & = & D_{c}K_{a}^{c}-D_{a}\left(K+%
\mathcal{K}\right)+\mathcal{K}_{a}L+\mathcal{L}_{\mathbf{l}}\mathcal{K}%
_{a}+M^{-1}K_{a}^{c}D_{c}M-M^{-1}\mathcal{K}D_{a}M\;\;,\label{gnG}%
\\
g_{a}^{c}l^{d}\widetilde{G}_{cd} & = & D_{c}L_{a}^{c}-D_{a}\left(L-%
\mathcal{L}\right)+\mathcal{K}_{a}K+\mathcal{L}_{\mathbf{n}%
}K_{a}+N^{-1}L_{a}^{c}D_{c}N+N^{-1}\mathcal{L}D_{a}N\;\;,\label{glG}%
\\
n^{a}n^{b}\widetilde{G}_{ab} & = & \frac{1}{2}%
\left(R-K_{ab}K^{ab}+K^{2}+3L_{ab}L^{ab}-L^{2}\right)+\mathcal{K}K-\mathcal{K%
}_{a}\mathcal{K}^{a}-g^{ab}\pounds_{\mathbf{l}}L_{ab}-M^{-1}D_{a}D^{a}M\;\;,%
\label{nnG}\\
l^{a}l^{b}\widetilde{G}_{ab} & = & -\frac{1}{2}%
\left(R+L_{ab}L^{ab}-L^{2}-3K_{ab}K^{ab}+K^{2}\right)-\mathcal{L}L+\mathcal{K%
}_{a}\mathcal{K}^{a}-g^{ab}\pounds_{\mathbf{n}}K_{ab}+N^{-1}D_{a}D^{a}N\;\;,%
\label{llG}\\
n^{a}l^{b}\widetilde{G}_{ab} & = & D_{a}\mathcal{K}%
^{a}-g^{ab}\pounds_{\mathbf{l}}K_{ab}+K_{ab}L^{ab}+\mathcal{K}L+M^{-1}%
\mathcal{K}^{a}D_{a}M\;\;.\label{nlG}
\end{eqnarray}
\end{subequations}
The last three
equations agree with Eqs. (D2d)-(D2f) of Paper {\textbf I}. From among them (\ref{%
nnG}) correctly yields the
Hamiltonian constraint, Eq. (\ref{constraintsVham}). \end{widetext}

\section{The derivation of the Lanczos equation}

We have demonstrated in Section V. that integration of the dynamical
equations and of the constraints across the brane gives the projections of
the Lanczos equation in the momentum phase space. In the velocity phase
space these projections were obtained directly by projections of the full
Lanczos equation, as Eqs. (53), (54), and (55) of Paper \textbf{I }(these
are however valid only for $s=3$). Equivalently, we can integrate the
dynamical equations (67a), (67c) and the projection (D2e) of the Ricci
tensor (i.e., the diffeomorphism constraint given in terms of velocity
instead of momentum) of Paper \textbf{I} in order to obtain in full
generality (for generic $s$) the projections of the Lanczos equation in the
velocity phase-space. With this we fully establish the commutativity of the
variation principle and the geometrical decomposition of the quantities, to
be discussed in more detail in Appendix C.

In order to carry out this program, we apply the regularization procedure
described in Section V. for the evolution equations of $K_{ab}$ and $%
\mathcal{K}$, derived in Paper \textbf{I}, which contain the following
projections of $\widetilde{R}_{ab}$. 
\begin{eqnarray}
g_{a}^{c}{}g_{b}^{d}\widetilde{R}_{cd}\! &=&\!\widetilde{\kappa }^{2}\!\left[
\frac{2\widetilde{\Lambda }+\left( \rho +\lambda \right) \delta \left( \chi
\right) }{s}g_{ab}+\Pi _{ab}\delta \left( \chi \right) \right] ~, \\
l^{a}l^{b}\widetilde{R}_{ab}\! &=&\!\widetilde{\kappa }^{2}\!\left[ \frac{2%
\widetilde{\Lambda }+\left[ \rho -sp+\left( s+1\right) \lambda \right]
\delta \left( \chi \right) }{s}\right] ~,
\end{eqnarray}%
The first and third of the Eqs. (67) of Paper \textbf{I} can be rewritten
conveniently as 
\begin{subequations}
\label{Kevol}
\begin{eqnarray}
\frac{\partial }{\partial t}K_{ab} &=&\widetilde{\kappa }^{2}N\left[ \frac{%
\left( \rho +\lambda \right) }{s}g_{ab}+\Pi _{ab}\right] \delta \left( \chi
\right)   \notag \\
&&+\frac{\partial }{\partial \chi }\left( \frac{N}{M}L_{ab}\right) +A_{ab}~,
\\
\frac{\partial }{\partial t}\mathcal{K} &=&\widetilde{\kappa }^{2}N\frac{%
\left[ \rho -sp+\left( s+1\right) \lambda \right] }{s}\delta \left( \chi
\right)   \notag \\
&&+\frac{\partial }{\partial \chi }\left[ \frac{N}{M}\left( L-\mathcal{L}%
\right) \right] +A~,
\end{eqnarray}%
where in $A_{ab}$ and $A$ we have collected only \textit{finite} terms: 
\end{subequations}
\begin{subequations}
\label{A}
\begin{eqnarray}
A_{ab} &=&N\Biggl[\frac{2}{s}\widetilde{\kappa }^{2}\widetilde{\Lambda }%
g_{ab}-R_{ab}+L_{ab}\left( L-\mathcal{L}\right) -2L_{ac}L_{b}^{c}  \notag \\
&&-K_{ab}\left( K+\mathcal{K}\right) +2K_{ac}K_{b}^{c}+2\mathcal{K}_{a}%
\mathcal{K}_{b}  \notag \\
&&+\frac{1}{M}\left( D_{b}D_{a}M-M^{c}D_{c}L_{ab}-2L_{c(a}D_{b)}M^{c}\right) %
\Biggr]  \notag \\
&&+D_{b}D_{a}N+N^{c}D_{c}K_{ab}+2K_{c(a}D_{b)}N^{c}  \notag \\
&&-L_{ab}\frac{\partial }{\partial \chi }\left( \frac{N}{M}\right) \ , \\
A &=&N\Biggl\{\frac{2}{s}\widetilde{\kappa }^{2}\widetilde{\Lambda }%
-L_{ab}L^{ab}+\mathcal{L}^{2}-2\mathcal{K}_{a}\mathcal{K}^{a}-\mathcal{K}%
\left( K+\mathcal{K}\right)   \notag \\
&&-\frac{1}{M}M^{a}D_{a}\left( L-\mathcal{L}\right) +\frac{D_{a}D^{a}M}{M}%
\Biggr\}  \notag \\
&&+\frac{D^{a}M}{M}D_{a}N+N^{a}D_{a}\mathcal{K}  \notag \\
&&-\left( L-\mathcal{L}\right) \frac{\partial }{\partial \chi }\left( \frac{N%
}{M}\right) \ .
\end{eqnarray}%
The integration across a finite coordinate distance containing the brane
(for example from $-\chi _{0}$ to $\chi _{0}$) of Eqs. (\ref{Kevol}) and the
subsequent limit $\chi _{0}\rightarrow 0$ gives 
\end{subequations}
\begin{subequations}
\label{Kevolint}
\begin{eqnarray}
-\widetilde{\kappa }^{2}\frac{N}{M}\left[ \frac{\left( \rho +\lambda \right) 
}{s}g_{ab}+\Pi _{ab}\right] \! &=&\!\Delta \left( \frac{N}{M}L_{ab}\right) ~,
\label{K1} \\
-\widetilde{\kappa }^{2}\frac{N}{M}\left[ \frac{\rho -sp+\left( s+1\right)
\lambda }{s}\right] \! &=&\!\Delta \left[ \frac{N}{M}\left( L-\mathcal{L}%
\right) \right] ~.  \label{K2}
\end{eqnarray}%
(In deriving Eqs. (\ref{Kevolint}) we have employed that time-derivatives
and integration over $\chi $ commute.) With the remark that neither $M$, nor 
$N$ are discontinuous across the brane, but the extrinsic curvature $L_{ab}$
is (as it depends on the brane embedding into the bulk on each side), from
Eq. (\ref{K1}) we obtain the jump of $L_{ab}$ as 
\end{subequations}
\begin{equation}
\Delta L_{ab}=-\widetilde{\kappa }^{2}\left[ \frac{\left( \rho +\lambda
\right) }{s}g_{ab}+\Pi _{ab}\right] ~.  \label{Lanczos_tensor}
\end{equation}%
This is the \textit{tensorial projection of the Lanczos equation}, Eq. (53)
of Paper \textbf{I}, valid for generic $s$. The trace of the right hand side
of this equation is 
\begin{equation}
\Delta L=-\widetilde{\kappa }^{2}\left( \rho +\lambda \right) ~.
\end{equation}%
With this, Eq. (\ref{K2}) implies%
\begin{equation}
\Delta \mathcal{L}=\widetilde{\kappa }^{2}\frac{\left( 1-s\right) \rho
-sp+\lambda }{s}~,  \label{Lanczos_scalar}
\end{equation}%
which is the \textit{scalar projection of the Lanczos equation}, Eq. (55) of
Paper \textbf{I}, valid for generic $s$.

The vectorial projection of the Lanczos equation does not emerge in a
similar way. Indeed, the remaining evolution equation for $\mathcal{K}_{a}$
[the second Eq. (67) of Paper \textbf{I}], is 
\begin{eqnarray}
\frac{\partial }{\partial t}\mathcal{K}_{a} &=&N\left[ -D^{b}L_{ab}+D_{a}%
\left( L-\mathcal{L}\right) -K\mathcal{K}_{a}\right] +N^{b}D_{b}\mathcal{K}%
^{a}  \notag \\
&&-\left( L_{a}^{b}+\mathcal{L}\delta _{a}^{b}\right) D_{b}N+\mathcal{K}%
_{b}D_{a}N^{b}\ .
\end{eqnarray}%
As all terms are regular, integration over the range $\left( -\chi _{0}\text{%
, }\chi _{0}\right) $ and the limit $\chi _{0}\rightarrow 0$ would give
nothing but the identity $0=0$.

The vectorial projection of the Lanczos equation can be instead obtained
from Eq. (D2e) of Paper \textbf{I}, which contains the projection $%
g_{a}^{c}{}n^{d}{}\widetilde{R}_{cd}$, given by the bulk Einstein equation:%
\begin{equation*}
g_{a}^{c}{}n^{d}\widetilde{R}_{cd}=\widetilde{\kappa }^{2}\left[
g_{a}^{c}{}n^{d}\widetilde{\Pi }_{cd}-Q_{a}\delta \left( \chi \right) \right]
\ .
\end{equation*}%
Written in terms of $\chi $-derivatives, Eq. (D2e) of Paper \textbf{I}
becomes%
\begin{gather}
0=\widetilde{\kappa }^{2}Q_{a}\delta \left( \chi \right) +\frac{1}{M}\frac{%
\partial }{\partial \chi }\mathcal{K}_{a}+A_{a} \\
A_{a}=-\widetilde{\kappa }^{2}g_{a}^{c}{}n^{d}\widetilde{\Pi }%
_{cd}+D_{c}K_{a}^{c}-D_{a}\left( K+\mathcal{K}\right) +\mathcal{K}_{a}L 
\notag \\
+\frac{1}{M}\left( K_{a}^{i}D_{i}M-\mathcal{K}D_{a}M+\mathcal{K}%
^{c}D_{c}M_{a}-M^{c}D_{c}\mathcal{K}_{a}\right) \ ,
\end{gather}%
where $A_{a}$ represents the collection of finite terms. Integration over an
infinitesimal range $\left( -\chi _{0}\text{, }\chi _{0}\right) $ and taking
the limit $\chi _{0}\rightarrow 0$ gives%
\begin{equation}
\Delta \mathcal{K}_{a}=-\widetilde{\kappa }^{2}Q_{a}\ .
\label{Lanczos_vector}
\end{equation}%
This is the \textit{vectorial projection of the Lanczos equation}, Eq. (54)
of Paper \textbf{I}.

\section{Dynamics regained from the $(s+1)+1$ ADM decomposition}

In this Appendix we regain the full $\left( s+1+1\right) $-break-up of the $%
\left( s+2\right) $-dimensional equations by a step-by-step method, by
further splitting of the $\left( (s+1)+1\right) $-dimensional ADM
decomposition of the bulk. First we foliate the $\left( s+2\right) $%
-dimensional space-time $\mathcal{B}$ with the $\left( s+1\right) $-spaces $%
\mathcal{S}_{t}$. Then we derive the equations of motion together with the
constraints from the corresponding variation principle. Finally we embed the 
$s$-spaces $\Sigma _{t\chi }$ in the hypersurfaces $\mathcal{S}_{t}$ and
perform a second split-up with respect to the extra dimension of the
geometrodynamics.

We begin by briefly presenting the standard $(s+1)+1$ ADM decomposition of
the vacuum Lagrangian. This yields: 
\begin{equation}
\mathcal{L}^{G}=N\sqrt{\widehat{g}}(\widehat{R}-\widehat{K}^{2}+\widehat{K}%
_{ab}\widehat{K}^{ab})-2\widetilde{\nabla }_{a}[N\sqrt{\widehat{g}}(\alpha
^{a}-\widehat{K}n^{a})]\;,  \label{L4}
\end{equation}%
with the dynamical variables $\widehat{g}_{ab}$ and $\widehat{K}_{ab}$,
representing the first and second fundamental forms of $\mathcal{S}_{t}$,
respectively. Here $\widehat{R}$ is the intrinsic curvature scalar of $%
\mathcal{S}_{t}$ and $\widehat{K}=\widehat{g}^{ab}\widehat{K}_{ab}$ is the
trace of $\widehat{K}_{ab}$. This equation can be rewritten in the form%
\begin{eqnarray}
\mathcal{L}^{G} &=&\sqrt{\widehat{g}}\left( \widehat{\pounds }_{\mathbf{t}}%
\widehat{K}+\widehat{g}^{ab}\widehat{\pounds }_{\mathbf{t}}\widehat{K}%
_{ab}\right) +N\widehat{\mathcal{H}}_{\bot }^{G}  \notag \\
&&+N_{a}\widehat{\mathcal{H}}_{a}^{G}-2\sqrt{\widehat{g}}\widehat{D}_{a}(%
\widehat{D}^{a}N+N_{b}\widehat{K}^{ab})]
\end{eqnarray}%
containing the pure spatial projections $\widehat{\pounds }_{\mathbf{t}}%
\widehat{K}_{ab}=\widehat{g}^{c}{}_{a}\widehat{g}^{d}{}_{b}\widetilde{%
\pounds }_{\mathbf{t}}\widehat{K}_{cd}$ and $\widehat{D}_{a}\widehat{K}_{bc}=%
\widehat{g}^{d}{}_{a}\widehat{g}^{e}{}_{b}\widehat{g}^{f}{}_{c}\widetilde{%
\nabla }_{d}\widehat{K}_{ef}$ of the $\left( s+2\right) $-dimensional
Lie-derivative $\widetilde{\pounds }_{\mathbf{t}}\widehat{K}_{cd}$ and
covariant derivative $\widetilde{\nabla }_{d}\widehat{K}_{ef}$,
respectively. We can also identify the cofactors of the Lagrange multipliers 
$N$ and $N^{a}$ as the super-Hamiltonian and super-momentum constraints 
\begin{subequations}
\begin{eqnarray}
\widehat{\mathcal{H}}_{\bot }^{G}\!\! &=&\!\!-2\sqrt{\widehat{g}}n^{a}n^{b}%
\widetilde{G}_{ab}\!=\!\sqrt{\widehat{g}}\left( \widehat{R}+\widehat{K}%
^{2}\!-\!\widehat{K}_{ab}\widehat{K}^{ab}\right) \;, \\
\widehat{\mathcal{H}}_{a}^{G}\!\! &=&\!\!-2\sqrt{\widehat{g}}\widehat{g}%
^{b}{}_{a}n^{c}\widetilde{G}_{bc}\!=\!2\sqrt{\widehat{g}}\widehat{D}_{b}(%
\widehat{K}^{ab}-\widehat{K}\widehat{g}^{ab})\;.
\end{eqnarray}%
By introducing the $\left( s+1\right) $-dimensional momentum 
\end{subequations}
\begin{equation}
\widehat{\pi }^{ab}=\frac{\partial L^{G}}{\partial (\widehat{\pounds }_{%
\mathbf{t}}\widehat{g}_{ab})}=\sqrt{\widehat{g}}(\widehat{K}^{ab}-\widehat{g}%
^{ab}\widehat{K})\;,
\end{equation}%
the vacuum action $S^{G}=\int d^{s+2}x\sqrt{-\widetilde{g}}L^{G}$ can be
cast into the form 
\begin{eqnarray}
&&S^{G}[\widehat{g}_{ab},\widehat{\pi }^{ab};\widehat{N}^{a},N]  \notag \\
&&\qquad =\int dt\int d^{s+1}x(\widehat{\pi }^{ab}\widehat{\pounds }_{%
\mathbf{t}}\widehat{g}_{ab}-N\widehat{\mathcal{H}}_{\bot }^{G}-N^{a}\widehat{%
\mathcal{H}}_{a}^{G})  \notag \\
&&\qquad \qquad -2\int d^{s+2}x\sqrt{-\widetilde{g}}\widetilde{\nabla }%
_{a}(\alpha ^{a}-\widehat{K}n^{a})  \notag \\
&&\qquad \qquad -2\int dt\int d^{s+1}x\widehat{D}_{a}(N_{b}\widehat{\pi }%
^{ab})\ ,  \label{act4}
\end{eqnarray}%
and the constraints transform to 
\begin{subequations}
\label{hatH}
\begin{eqnarray}
\widehat{\mathcal{H}}_{\bot }^{G}(\widehat{x};\widehat{g}_{ab},\widehat{\pi }%
^{ab}] &=&-\sqrt{\widehat{g}}\widehat{R}  \notag \\
&&+\frac{1}{\sqrt{\widehat{g}}}\left( \widehat{\pi }^{ab}\widehat{\pi }_{ab}-%
\frac{1}{s}\widehat{\pi }^{2}\right) \ , \\
\widehat{\mathcal{H}}_{a}(\widehat{x};\widehat{g}_{ab},\widehat{\pi }^{ab}]
&=&-2\widehat{D}^{b}\widehat{\pi }_{ab}
\end{eqnarray}%
for any point $\widehat{x}\in S_{t}$. Extremizing the action (\ref{act4})
with respect to the canonical variables $\widehat{g}_{ab}$ and $\widehat{\pi 
}^{ab}$ yields the field equations 
\end{subequations}
\begin{eqnarray}
\widehat{\pounds }_{\mathbf{t}}\widehat{g}_{ab} &=&2N\widehat{g}%
^{-1/2}\left( \widehat{\pi }_{ab}-\frac{1}{n}\widehat{\pi }\widehat{g}%
_{ab}\right) +\widehat{\pounds }_{\mathbf{N}}\widehat{g}_{ab}\;\;,
\label{hatLhatg} \\
\widehat{\pounds }_{\mathbf{t}}\widehat{\pi }^{ab} &=&-2N\widehat{g}%
^{-1/2}\left( \widehat{\pi }^{a}{}_{c}\widehat{\pi }^{bc}-\frac{1}{n}%
\widehat{\pi }\widehat{\pi }^{ab}\right)   \notag \\
&&+\frac{1}{2}N\widehat{g}^{-1/2}\left( \widehat{\pi }_{cd}\widehat{\pi }%
^{cd}-\frac{1}{n}\widehat{\pi }^{2}\right) \widehat{g}^{ab}+\widehat{\pounds 
}_{\mathbf{N}}\widehat{\pi }^{ab}  \notag \\
&&-N\widehat{g}^{1/2}\left( \widehat{R}^{ab}-\frac{1}{2}\widehat{R}\widehat{g%
}^{ab}\right)   \notag \\
&&+\widehat{g}^{1/2}(\widehat{D}^{a}\widehat{D}^{b}N-\widehat{g}^{ab}%
\widehat{D}^{c}\widehat{D}_{c}N)\;,  \label{hatLhatpi}
\end{eqnarray}%
whereas extremizing the action with respect to the Lagrange multipliers $N$
and $\widehat{N}^{a}$ yields the super-Hamiltonian and the super-momentum
constraints of vacuum gravity: $\widehat{\mathcal{H}}_{\bot }^{G}=0$ and $%
\widehat{\mathcal{H}}_{a}=0$.

The dynamical equations (\ref{hatLhatg}) and (\ref{hatLhatpi}) can be
equally interpreted as the time evolution of the canonical variables,
generated by the smeared constraints $\widehat{H}^{G}[N]=\widehat{H}_{\bot
}^{G}[N]+\widehat{H}_{a}^{G}[N^{a}]$: 
\begin{subequations}
\begin{eqnarray}
\widehat{\pounds }_{\mathbf{t}}\widehat{g}_{ab} &=&\{\widehat{g}_{ab},%
\widehat{H}^{G}[N]\}\;, \\
\widehat{\pounds }_{\mathbf{t}}\widehat{\pi }^{ab} &=&\{\widehat{\pi }^{ab},%
\widehat{H}^{G}[N]\}~,
\end{eqnarray}%
where the Poisson bracket is defined by 
\end{subequations}
\begin{eqnarray}
\{f(\widehat{x}),h(\widehat{x}^{\prime })\} &=&\int_{S_{t}}d^{s+1}\widehat{x}%
^{\prime \prime }\frac{\delta f(\widehat{x})}{\delta \widehat{g}_{ab}(%
\widehat{x}^{\prime \prime })}\frac{\delta h(\widehat{x}^{\prime })}{\delta 
\widehat{\pi }^{ab}(\widehat{x}^{\prime \prime })}  \notag \\
&-&\int_{S_{t}}d^{s+1}\widehat{x}^{\prime \prime }\frac{\delta f(\widehat{x})%
}{\delta \widehat{\pi }^{ab}(\widehat{x}^{\prime \prime })}\frac{\delta h(%
\widehat{x}^{\prime })}{\delta \widehat{g}_{ab}(\widehat{x}^{\prime \prime })%
}
\end{eqnarray}%
for any function $f(\widehat{x};\widehat{g}_{ab},\widehat{\pi }^{ab}]$ and $%
g(\widehat{x};\widehat{g}_{ab},\widehat{\pi }^{ab}]$. The Poisson brackets
of the constraints give the Dirac algebra 
\begin{subequations}
\begin{eqnarray}
\{\widehat{\mathcal{H}}_{\bot }^{G}(\widehat{x}),\widehat{\mathcal{H}}_{\bot
}^{G}(\widehat{x}^{\prime })\} &=&\widehat{g}^{ab}(\widehat{x})\widehat{%
\mathcal{H}}_{a}^{G}\left( \widehat{x}\right) \delta ,_{b}\left( \widehat{x},%
\widehat{x}^{\prime }\right)  \notag \\
&&-\left( \widehat{x}\leftrightarrow \widehat{x}^{\prime }\right) \ , \\
\{\widehat{\mathcal{H}}_{\bot }^{G}(\widehat{x}),\widehat{\mathcal{H}}%
_{a}^{G}(\widehat{x}^{\prime })\} &=&\widehat{\mathcal{H}}^{G}\bot (\widehat{%
x})\delta ,_{a}\left( \widehat{x},\widehat{x}^{\prime }\right)  \notag \\
&&+\widehat{\mathcal{H}}_{\bot ,a}^{G}(\widehat{x})\delta \left( \widehat{x},%
\widehat{x}^{\prime }\right) \ , \\
\{\widehat{\mathcal{H}}_{a}^{G}(\widehat{x}),\widehat{\mathcal{H}}_{b}^{G}(%
\widehat{x}^{\prime })\} &=&\widehat{\mathcal{H}}_{b}^{G}(\widehat{x})\delta
,_{a}\left( \widehat{x},\widehat{x}^{\prime }\right)  \notag \\
&&-\left( a\widehat{x}\leftrightarrow b\widehat{x}^{\prime }\right) \ .
\end{eqnarray}%
The dynamical equations (\ref{hatLhatg}) and (\ref{hatLhatpi}) together with
the constraints (\ref{hatH}) provide the full geometrodynamics of the vacuum
gravity. Any matter field can be equally $\left( (s+1)+1\right) $%
-dimensionally decomposed and coupled to gravity.

The next step necessary in order to recover the $\left( s+1+1\right) $%
-dimensional decomposition of geometrodynamics is a further split. We can
interpret the momenta $\pi ^{ab},\ p_{a}$ and $p$ introduced in the main
text as the projections of $\widehat{\pi }^{ab}$: 
\end{subequations}
\begin{equation}
\widehat{\pi }^{ab}=\pi ^{ab}+Ml^{(a}p^{b)}+\frac{M}{2}l^{a}l^{b}p\ .
\label{d2}
\end{equation}%
By inserting Eq. (\ref{d2}) into the action (\ref{act4}), we regain the
fully decomposed action (\ref{actioncan}) with the constraints (\ref{constrG}%
). Similarly, the dynamical equations (\ref{hatLhatg}) and (\ref{hatLhatpi})
can be split by applying the formula (\ref{d2}). The only non-trivial step
is the decomposition of the Lie-derivatives. For this we employ: 
\begin{eqnarray}
&&(\widehat{\pounds }_{\mathbf{t}}-\widehat{\pounds }_{\mathbf{N}})\widehat{%
\pi }^{ab}  \notag \\
&=&\left( \frac{\partial }{\partial t}-\pounds _{\mathbf{N}}\right) \pi
^{ab}+Ml^{(a}\left( \frac{\partial }{\partial t}-\pounds _{\mathbf{N}%
}\right) p^{b)}  \notag \\
&&+\frac{M}{2}l^{a}l^{b}\left( \frac{\partial }{\partial t}-\pounds _{%
\mathbf{N}}\right) p-\frac{NM}{\sqrt{g}}(p^{a}p^{b}+pl^{(a}p^{b)})  \notag \\
&&-\frac{N}{2s\sqrt{g}}\left( \frac{s-1}{2}Mp^{2}-\pi p\right)
l^{a}l^{b}\;\;.
\end{eqnarray}%
With this we regain the fully decomposed dynamical equations (\ref{dotgab})-(%
\ref{dotp}). From Eqs. (\ref{d2}) and (\ref{hatH}) we can also obtain the
constraints (\ref{constrG}), by using the twice contracted Gauss equation
for the hypersurface $\Sigma _{t\chi }$ of $\mathcal{S}_{t}$: 
\begin{equation}
\widehat{R}=R+L^{2}-L_{ab}L^{ab}+2\widehat{D}_{a}(l^{c}\widehat{D}%
_{c}l^{a}-Ll^{a})\ .  \label{d4}
\end{equation}

We have thus shown that the variation principle and the $\left( s+1+1\right) 
$-decomposition commute, i.e. no matter which order we apply the
decomposition of the bulk and the extremization of the action, the result is
the same.

\end{document}